\documentclass[11pt]{article}
\usepackage{epsfig}
\usepackage{axodraw}
\newcommand{\sect}[1]{ \section{#1} \setcounter{equation}{0} }

\newcommand{\Dslash}{D \! \! \! \! /}

\newcommand{\third}{\mbox{\small{$\frac{1}{3}$}}} 
 
\newcommand{\pitwo}{\mbox{\small{$\frac{\pi}{2}$}}} 
\newcommand{\pisix}{\mbox{\small{$\frac{\pi}{6}$}}} 
\newcommand{\MSbar}{\overline{\mbox{MS}}} 
\newcommand{\MSbars}{\overline{\mbox{\footnotesize{MS}}}} 
\newcommand{\MOMg}{\mbox{MOMggg}}
\newcommand{\MOMgs}{\mbox{\footnotesize{MOMggg}}}
\newcommand{\MOMh}{\mbox{MOMh}}
\newcommand{\MOMhs}{\mbox{\footnotesize{MOMh}}}
\newcommand{\MOMq}{\mbox{MOMq}}
\newcommand{\MOMqs}{\mbox{\footnotesize{MOMq}}}
\newcommand{\MOMi}{\mbox{MOMi}}
\newcommand{\MOMis}{\mbox{\footnotesize{MOMi}}}
\newcommand{\Nc}{N_{\!c}}
\newcommand{\Nf}{N_{\!f}}
\newcommand{\NF}{N_{\!F}}
\newcommand{\NA}{N_{\!A}}
\newcommand{\Nda}{N^d_{\!A}}
\newcommand{\Noda}{N^o_{\!A}}

\begin{document}
\title{Maximal Abelian and Curci-Ferrari gauges in momentum subtraction at 
three loops}
\author{J.M. Bell \& J.A. Gracey, \\ Theoretical Physics Division, \\ 
Department of Mathematical Sciences, \\ University of Liverpool, \\ P.O. Box 
147, \\ Liverpool, \\ L69 3BX, \\ United Kingdom.} 
\date{} 
\maketitle 

\vspace{5cm} 
\noindent 
{\bf Abstract.} The vertex structure of QCD fixed in the maximal Abelian gauge
(MAG) and Curci-Ferrari gauge is analysed at two loops at the fully symmetric 
point for the $3$-point functions corresponding to the three momentum 
subtraction (MOM) renormalization schemes. Consequently the {\em three} loop 
renormalization group functions are determined for each of these three schemes 
in each gauge using properties of the renormalization group equation. 

\vspace{-16.0cm}
\hspace{13cm}
{\bf LTH 1062}

\newpage

\sect{Introduction.}

Four dimensional gauge theories are of interest due to their description of the
elementary quanta of nature. For instance, Quantum Chromodynamics (QCD) 
underpins our theoretical understanding of the strong nuclear force where the
basic fields are quarks and gluons. These behave as effectively free particles
but only in the high energy limit due to asymptotic freedom, \cite{1,2}. At low
energies quarks and gluons are actually confined and do not exist in nature as 
free particles. The reason why this is the case is one of the major problems in
quantum field theory. Various ideas as to the specific confinement mechanism 
have been proposed. One which is popular is the dual superconductor ideas of 
\cite{3,4,5,6}. There the colour electric flux is restricted by the Meissner 
effect of superconductivity. The role of the Cooper pair condensation of 
superconductivity is played by the condensation of colour magnetic monopoles in
the non-Abelian gauge theory case. While this picture parallels 
superconductivity it is difficult to access the underlying dynamics 
practically. One approach is to use an Abelian projection, \cite{3,4,6,7}, 
which appears to give insight into low energy properties of colour confinement.
For a non-Abelian gauge theory the maximal Abelian sector derives from the 
centralizer of the colour group. Clearly as such phenomena lie deep within the 
infrared non-perturbative regime they can only theoretically be examined by 
lattice regularized gauge theory or Schwinger-Dyson techniques. Underlying both 
approaches is the need to isolate the Abelian degrees of freedom to effect a 
study of this monopole model. Linked to this in a Lagrangian analysis is the 
need to fix the gauge. Ordinarily one computes in linear covariant gauges such 
as the Landau gauge. However, this does not have the feature of readily 
distinguishing the colour group centralizer which is related to the Abelian 
projection. A more appropriate gauge is the maximal Abelian gauge (MAG) 
introduced in \cite{6,8,9}. There the gluons in the sectors delineated by 
whether their associated generators totally commute among themselves or not are
gauge fixed differently. For instance, the diagonal gluons, corresponding to 
the subgroup of generators which totally commute, are fixed in the Landau gauge
but the remaining off-diagonal gluons are gauge fixed by a different fixing 
criterion, \cite{6,8,9}. Ultimately a covariant but nonlinear gauge fixing 
emerges but in a way which produces a renormalizable Lagrangian. The 
renormalizability has been established in \cite{10,11,12,13,14,15}. Properties 
of the MAG have been studied in various contexts. An interesting recent lattice
study, for instance, was in \cite{16} where the effect of the diagonal gluons 
on the inter-quark static potential was examined. In particular within the 
theoretical setup it was possible to identify the contributions made by the 
diagonal gluons to the potential. Excluding these it was demonstrated, 
\cite{16}, that the linearly rising potential collapsed indicating that the 
Abelian sector was effectively responsible for quark confinement.

As such infrared lattice studies are important, from a more theoretical point 
of view the Lagrangian field theory focus is concerned with the $2$- and 
$3$-point functions of QCD. The low energy behaviour of the former Green's
function diverges from the canonical structure of a fundamental particle, 
\cite{17,18,19,20,21,22,23,24,25,26,27}, while the latter are studied to assist
with building models of hadrons for instance. Therefore Schwinger-Dyson studies
have centred on the properties of these Green's functions and particularly in 
the MAG, \cite{28,29,30}. Lattice studies in the MAG can be found, for 
instance, in \cite{31,32,33,34,35,36}. Related to this is the need to 
ultimately overlap with conventional perturbative analyses. Lattice 
measurements of vertex functions and Schwinger-Dyson studies have to 
consistently match onto known perturbation theory. This was partly the 
motivation to a previous study, \cite{37}, where the structure of the $3$-point
vertices of the MAG fixed QCD Lagrangian were evaluated at one loop at the 
completely symmetric point. One reason for examining these functions at this 
point is that this momentum configuration is non-exceptional. So there are no 
infrared issues unlike the case where the external momentum of one external leg
is nullified. Although the latter is a much more widely studied configuration. 
As a corollary of the one loop analysis of \cite{37} the momentum subtraction 
(MOM) scheme renormalization group functions of the MAG were derived at 
{\em two} loops from properties of the renormalization group equation. In this 
context as well as the need for previous matching of vertex functions for 
lattice and Schwinger-Dyson techniques it is the purpose of this article to 
extend the results of \cite{37} to the next loop order. By this we mean the 
full structure of the two loop vertex functions corresponding to the MOM 
schemes, introduced by Celmaster and Gonsalves in \cite{38,39}, and hence 
deduce the {\em three} loop renormalization group functions. A separate but 
parallel motivation concerns the relation of the MAG to another nonlinear 
covariant gauge fixing. This is the Curci-Ferrari gauge which involves a 
quartic ghost interaction unlike the canonical linear covariant gauge. This
gauge fixing was introduced in \cite{40} in part to study massive vector bosons
without symmetry breaking. A mass term for the gluons and ghosts could be 
introduced in a Becchi-Rouet-Stora-Tyutin (BRST) invariant way. In relation to 
the MAG the Curci-Ferrari gauge fixed Lagrangian emerges in a specific limit. 
This is the case when the diagonal fields are formally removed from the MAG 
Lagrangian. Aside from the diagonal gluons this includes the associated 
diagonal ghosts which together with the off-diagonal ghosts derive from the 
Faddeev-Popov technique, \cite{8,9,10,11,12,13,14,15}. Therefore, the emerging 
MOM renormalization group functions in the same limit must agree precisely with 
those of the Curci-Ferrari gauge. These are also computed directly here as an 
important independent check on the MAG results. One interesting feature of the 
MAG, which may have bearing on the infrared properties alluded to already, is 
that the diagonal gluons appear to play a similar role to the background gluons
of the background field gauge of \cite{41,42,43,44,45,46}. This is because the 
anomalous dimension of the diagonal gluon is precisely proportional to the 
$\beta$-function. This has been established to all orders in perturbation 
theory from the Slavnov-Taylor identities constructed during the algebraic 
renormalization proof of the renormalizability of the MAG, \cite{15}.

The paper is organized as follows. In the two subsequent sections we introduce 
and review all the relevant renormalization background and computation methods 
required for our study of MAG and Curci-Ferrari gauges in QCD at the fully 
symmetric point of the $3$-point vertices. The results of the application of 
this formalism are given in section $4$. Finally, conclusions are presented in 
section $5$.

\sect{Background.}

In this section we record the relevant aspects of the MAG and its relation to
the Curci-Ferrari gauge for renormalization at the symmetric point in the
various MOM renormalization schemes of \cite{38,39} including the structure of 
the gauge fixed QCD Lagrangian in the MAG. The main ingredient for the MAG is 
that the diagonal gluons are treated differently from the off-diagonal ones. 
Therefore, in keeping with other work, \cite{15}, we write the group valued 
gluon field ${\cal A}_\mu$ as
\begin{equation}
{\cal A}_\mu ~\equiv~ A^A_\mu T^A ~=~ A^a_\mu T^a ~+~ A^i_\mu T^i
\end{equation}
where $T^A$ are the colour group generators. On notation we will use upper case
Roman letters, such as $A$, $B$, $C$ and $D$, for adjoint colour indices but 
lower case where the associated field is either in the centre or is 
off-diagonal. These are distinguished by using $a$, $b$, $c$ and so on for the 
off-diagonal fields except that $i$, $j$, $k$ and $l$ are reserved exclusively
and unambiguously for diagonal indices. Therefore, each set of indices run over
different ranges which are $1$~$\leq$~$A$~$\leq$~$\NA$, 
$1$~$\leq$~$a$~$\leq$~$\Noda$, and $1$~$\leq$~$i$~$\leq$~$\Nda$. (We use the 
notation of \cite{47} throughout.) Here $\NA$ is the dimension of the adjoint 
representation of the colour group, $\Nda$ is the dimension of the diagonal
subgroup with $\Noda$ being the dimension of the off-diagonal sector. Clearly, 
\begin{equation}
\Nda ~+~ \Noda ~=~ \NA ~.
\end{equation}
For reference, if the colour group is $SU(\Nc)$ then $\Nda$~$=$~$(\Nc-1)$ and
$\Noda$~$=$~$\Nc(\Nc-1)$. With this splitting of the gluons into separate 
sectors one has to reconsider the canonical group theory required to perform
the loop computations. The necessary relations can be established from the 
usual group identities such as the definition of the group Casimirs and Jacobi 
identities. For instance, 
\begin{equation}
\mbox{Tr} \left( T^a T^b \right) ~=~ T_F \delta^{ab} ~~,~~
\mbox{Tr} \left( T^a T^i \right) ~=~ 0 ~~,~~
\mbox{Tr} \left( T^i T^j \right) ~=~ T_F \delta^{ij} 
\label{Tfdef}
\end{equation}
follow from 
\begin{equation}
\mbox{Tr} \left( T^A T^B \right) ~=~ T_F \delta^{AB} ~.
\end{equation}
Equally by allowing the free indices in the Lie algebra
\begin{equation}
\left[ T^A , T^B \right] ~=~ i f^{ABC} T^C 
\end{equation}
to lie in the two sectors separately it is straightforward to deduce 
\begin{equation}
f^{ijk} ~=~ 0 ~~~,~~~ f^{ijc} ~=~ 0 
\end{equation}
whence 
\begin{equation}
\left[ T^a , T^j \right] ~=~ i f^{ajc} T^c ~. 
\end{equation}
Using these basic observations and the Casimir definitions 
\begin{equation}
f^{ACD} f^{BCD} ~=~ C_A \delta^{AB} ~~,~~ T^A T^A ~=~ C_F I 
\end{equation}
where the subscript in $C_A$ is not a summed index, one can deduce, \cite{47}, 
\begin{eqnarray}
T^i T^i &=& \frac{T_F}{\NF} \Nda I ~~,~~ 
T^a T^a ~=~ \left[ C_F ~-~ \frac{T_F}{\NF} \Nda \right] I \nonumber \\ 
C_A \delta^{ab} &=& f^{acd} f^{bcd} ~+~ 2 f^{acj} f^{bcj} ~~,~~ 
C_A \delta^{ij} ~=~ f^{icd} f^{jcd} ~~,~~ 
f^{abc} f^{abc} ~=~ \left[ \Noda - 2 \Nda \right] C_A \nonumber \\
f^{iab} f^{iab} &=& \Nda C_A ~~,~~
f^{acj} f^{bcj} ~=~ \frac{\Nda}{\Noda} C_A \delta^{ab} ~~,~~
f^{acd} f^{bcd} ~=~ \frac{[\Noda - 2 \Nda]}{\Noda} C_A \delta^{ab} ~.
\end{eqnarray}
In addition the Jacobi identity
\begin{equation}
0 ~=~ f^{ABE} f^{CDE} ~+~ f^{BCE} f^{ADE} ~+~ f^{CAE} f^{BDE} 
\end{equation}
implies
\begin{eqnarray}
f^{apq} f^{bpr} f^{cqr} &=& \frac{[\Noda - 3 \Nda]}{2\Noda} C_A f^{abc} ~~,~~
f^{apq} f^{bpi} f^{cqi} ~=~ \frac{\Nda}{2\Noda} C_A f^{abc}
\nonumber \\
f^{ipq} f^{bpr} f^{cqr} &=& \frac{[\Noda - 2 \Nda]}{2\Noda} C_A f^{ibc} ~~,~~
f^{ipq} f^{bpj} f^{cqj} ~=~ \frac{\Nda}{\Noda} C_A f^{ibc} 
\end{eqnarray}
where we note that here we briefly use $p$, $q$ and $r$ to denote off-diagonal
indices. All these relations were required for the evaluation of the two loop 
vertex functions.

As we will be considering two gauges in this article we note that the general
form of the QCD Lagrangian is
\begin{equation}
L ~=~ -~ \frac{1}{4} G^A_{\mu\nu} G^{A \, \mu\nu} ~+~ 
i \bar{\psi} \Dslash \psi ~+~ L_{\mbox{\footnotesize{gf}}} 
\end{equation}
with $\psi$ representing $\Nf$ flavours of massless quarks and 
\begin{equation}
L_{\mbox{\footnotesize{gf}}} ~=~ -~ \frac{1}{2\alpha} F[A_\mu^a]^2 ~-~ 
\frac{1}{2\bar{\alpha}} F[A_\mu^i]^2 ~+~ \bar{c}^A 
\left( \frac{\delta F[A_{U\,\mu}]}{\delta U} \right)^{AB} c^B
\end{equation}
where $F[A_\mu]$ is the functional of the gauge field whose explicit forms
define the different gauges, $A_{U\,\mu}$ is the gauge field under a general
gauge transformation $U$ and $c^A$ and $\bar{c}^A$ are the Faddeev-Popov 
ghosts. In our case we have
\begin{equation}
F[A^A_\mu] ~=~ \left\{
\begin{array}{ll}
\left( D^\mu A_\mu \right)^a ~+~ \frac{1}{2} \alpha b^a ~-~ 
\frac{1}{2} \alpha g f^{abi} \bar{c}^b c^i ~-~
\frac{1}{4} \alpha g f^{abc} \bar{c}^b c^c & \mbox{if} ~~ A ~=~ a \\
\frac{1}{\bar{\alpha}} \partial^\mu A^i_\mu & \mbox{if} ~~ A ~=~ i \\
\end{array}
\right.
\label{magfix}
\end{equation} 
for the MAG where the covariant derivative $D_\mu^{ab}$ acting on the 
off-diagonal sector is, for instance, 
\begin{equation}
\left( D_\mu A_\nu \right)^a ~=~ \partial_\mu A^a_\nu ~-~ g f^{abi} A^i_\mu
A^b_\nu ~~,~~ 
\left( D_\mu c \right)^a ~=~ \partial_\mu c^a ~-~ g f^{abi} A^i_\mu c^b
\end{equation}
and $\alpha$ is the gauge parameter for the off-diagonal sector. It is not to 
be confused with a similar parameter used in the canonical linear covariant 
gauge fixing. The other gauge fixing parameter, $\bar{\alpha}$, is the 
parameter associated with the diagonal sector. It is included here for 
completeness but throughout it will be set to zero, \cite{10,11,12,13,14,15}, 
so that the diagonal gluons are in the Landau gauge. In addition $c^i$ and 
$c^a$ are the Faddeev-Popov ghosts associated with the diagonal and 
off-diagonal sectors and $g$ is the coupling constant. For the Curci-Ferrari 
gauge
\begin{equation}
F[A^A_\mu] ~=~ \partial^\mu A^A_\mu ~+~ \frac{\alpha}{2} b^A ~-~ 
\frac{\alpha}{4} g f^{ABC} \bar{c}^B c^C 
\label{cffix}
\end{equation} 
with $\bar{\alpha}$~$=$~$\alpha$ and $b^A$ is the Nakanishi-Lautrup field for 
the respective gauges, \cite{15,40}. Although we have already eliminated $b^i$ 
in the MAG as the diagonal sector has a simple Abelian structure. From these 
functionals it is apparent that in the limit where the diagonal fields in the 
MAG are nullified then the Curci-Ferrari gauge fixing condition emerges if one 
identifies the off-diagonal indices with those of the full colour group. 

The first functional leads to the Lagrangian, \cite{15}, 
\begin{equation}
L^{\mbox{\footnotesize{MAG}}} ~=~ -~ 
\frac{1}{4} G^a_{\mu\nu} G^{a \, \mu\nu} ~-~
\frac{1}{4} G^i_{\mu\nu} G^{i \, \mu\nu} ~+~ i \bar{\psi} \Dslash \psi ~+~
L^{\mbox{\footnotesize{MAG}}}_{\mbox{\footnotesize{gf}}} ~. 
\end{equation}
The first two terms derive from the square of the field strength and their sum 
is gauge invariant. There is no cross term due to (\ref{Tfdef}). We have 
isolated the gauge fixing term  
$L^{\mbox{\footnotesize{MAG}}}_{\mbox{\footnotesize{gf}}}$. In a linear
covariant gauge the corresponding term contains the gauge fixing condition and
the consequent ghost Lagrangian. For the MAG the situation is the same but the
actual Lagrangian is more complicated since, \cite{10,11,12,13,14,15},
\begin{eqnarray}
L^{\mbox{\footnotesize{MAG}}}_{\mbox{\footnotesize{gf}}} &=& 
-~ \frac{1}{2\alpha} \left( \partial^\mu A^a_\mu \right)^2 
- \frac{1}{2\bar{\alpha}} \left( \partial^\mu A^i_\mu \right)^2 
+ \bar{c}^A \partial^\mu \partial_\mu c^A \nonumber \\
&& +~ g \left[ f^{abC} A^a_\mu \bar{c}^C \partial^\mu c^b
- \frac{1}{\alpha} f^{abk} \partial^\mu A^a_\mu A^b_\nu A^{k \, \nu}
- f^{abk} \partial^\mu A^a_\mu c^b \bar{c}^k
- \frac{1}{2} f^{abc} \partial^\mu A^a_\mu \bar{c}^b c^c 
\right. \nonumber \\
&& \left. ~~~~~~
- 2 f^{abk} A^k_\mu \bar{c}^a \partial^\mu \bar{c}^b
- f^{abk} \partial^\mu A^k_\mu \bar{c}^b c^c \right] \nonumber \\
&& +~ g^2 \left[ f_d^{acbd} A^a_\mu A^{b \, \mu} \bar{c}^c c^d
- \frac{1}{2\alpha} f_o^{akbl} A^a_\mu A^{b \, \mu} A^k_\nu A^{l \, \nu}
+ f_o^{adcj} A^a_\mu A^{j \, \mu} \bar{c}^c c^d \right.
\nonumber \\
&& \left. ~~~~~~~
- \frac{1}{2} f_o^{ajcd} A^a_\mu A^{j \, \mu} \bar{c}^c c^d
+ f_o^{ajcl} A^a_\mu A^{j \, \mu} \bar{c}^c c^l
+ f_o^{alcj} A^a_\mu A^{j \, \mu} \bar{c}^c c^l \right.
\nonumber \\
&& \left. ~~~~~~~
- f_o^{cjdi} A^i_\mu A^{j \, \mu} \bar{c}^c c^d
- \frac{\alpha}{4} f_d^{abcd} \bar{c}^a \bar{c}^b c^c c^d
- \frac{\alpha}{8} f_o^{abcd} \bar{c}^a \bar{c}^b c^c c^d
+ \frac{\alpha}{8} f_o^{acbd} \bar{c}^a \bar{c}^b c^c c^d \right. \nonumber \\
&& \left. ~~~~~~~
+ \frac{\alpha}{4} f^{albc} \bar{c}^a \bar{c}^b c^c c^l
- \frac{\alpha}{4} f_o^{albc} \bar{c}^a \bar{c}^b c^c c^l
+ \frac{\alpha}{2} f_o^{akbl} \bar{c}^a \bar{c}^b c^k c^l \right] 
\label{lagmag}
\end{eqnarray}
after eliminating $b^a$ where we have introduced the shorthand notation 
\begin{equation}
f_d^{ABCD} ~=~ f^{iAB} f^{iCD} ~~~,~~~
f_o^{ABCD} ~=~ f^{eAB} f^{eCD} 
\end{equation}
and 
\begin{equation}
f^{ABCD} ~=~ f_d^{ABCD} ~+~ f_o^{ABCD}
\end{equation}
for the quartic interaction terms. Hence the Jacobi identity is
\begin{equation}
f^{ABCD} ~+~ f^{ACDB} ~+~ f^{ADBC} ~=~ 0 ~.
\end{equation} 
As noted in \cite{47} the gauge fixed part of the MAG Lagrangian is generated 
automatically via a computer algebra routine from the BRST variation of the 
defining functional. This is to ensure that definitions and conventions are 
correctly implemented without error as well as to be confident that the 
resulting Feynman rules are derived correctly using symbolic manipulation. 
While the form of (\ref{lagmag}) is large we have endeavoured to condense the 
structure to save space. However, the nature of the MAG with the split in the 
colour group means that $L^{\mbox{\footnotesize{MAG}}}$ cannot be fully reduced
to a form which involves only the general indices $A$. For the Curci-Ferrari 
gauge using the second functional we have 
\begin{equation}
L^{\mbox{\footnotesize{CF}}} ~=~ -~ 
\frac{1}{4} G^A_{\mu\nu} G^{A \, \mu\nu} ~+~ i \bar{\psi} \Dslash \psi ~+~
L^{\mbox{\footnotesize{CF}}}_{\mbox{\footnotesize{gf}}} 
\end{equation}
with
\begin{eqnarray}
L^{\mbox{\footnotesize{CF}}}_{\mbox{\footnotesize{gf}}} &=& -~ 
\frac{1}{2\alpha} (\partial^\mu A^A_\mu)^2 ~-~
\bar{c}^A \left(\partial^\mu D_\mu c\right)^A \nonumber \\
&& -~ \frac{g}{2} f^{ABC} \partial^\mu A^A_\mu \, \bar{c}^B c^C ~+~
\frac{\alpha g^2}{8} f^{ABCD} \bar{c}^A c^B \bar{c}^C c^D ~. 
\label{lagcf}
\end{eqnarray}
While this is a more compact Lagrangian it is straightforward to check that it
is connected with (\ref{lagmag}) in the following way. Setting the diagonal 
gluon and ghost formally to zero in (\ref{lagmag}) then both Lagrangians are 
equivalent with the proviso that the adjoint group indices $A$ of (\ref{lagcf})
are equated with the off-diagonal ones of (\ref{lagmag}). In other words if one
takes the formal limit $\Nda$~$\rightarrow$~$0$ then the Curci-Ferrari gauge 
emerges from (\ref{lagmag}). This property was noted in \cite{47} for the three 
loop $\MSbar$ renormalization group functions but as indicated above, this is
also evident from the nature of the gauge fixing. We will exploit this 
observation later in our computations as a non-trivial check. Unlike the linear
covariant gauge both the MAG and Curci-Ferrari gauges have quartic ghost 
interactions but only the MAG has quartic gluon-ghost interactions.

Next it has been shown that both Lagrangians (\ref{lagmag}) and (\ref{lagcf}) 
are renormalizable, \cite{10,11,12,13,14,15,40,48,49,50,51,52,53}. However, 
from an algebraic renormalization analysis the general structure of the 
renormalization in the MAG has several subtleties. If we define the 
renormalization constants via the relationship from bare quantities, denoted by
${}_{\mbox{\footnotesize{o}}}$, to renormalized ones we have
\begin{eqnarray}
A^{a \, \mu}_{\mbox{\footnotesize{o}}} &=& \sqrt{Z_A} \, A^{a \, \mu} \,,\,
A^{i \, \mu}_{\mbox{\footnotesize{o}}} ~=~ \sqrt{Z_{A^i}} \, A^{i \, \mu} \,,\,
c^a_{\mbox{\footnotesize{o}}} ~=~ \sqrt{Z_c} \, c^a \,,\,
\bar{c}^a_{\mbox{\footnotesize{o}}} ~=~ \sqrt{Z_c} \, \bar{c}^a  \,,\,
c^i_{\mbox{\footnotesize{o}}} ~=~ \sqrt{Z_{c^i}} c^i \nonumber \\
\bar{c}^i_{\mbox{\footnotesize{o}}} &=& \frac{\bar{c}^i}{\sqrt{Z_{c^i}}} \,,\,
\psi_{\mbox{\footnotesize{o}}} ~=~ \sqrt{Z_\psi} \psi \,,\,
g_{\mbox{\footnotesize{o}}} ~=~ \mu^\epsilon Z_g \, g \,,\,
\alpha_{\mbox{\footnotesize{o}}} ~=~ Z^{-1}_\alpha Z_A \, \alpha \,,\,
\bar{\alpha}_{\mbox{\footnotesize{o}}} ~=~ Z^{-1}_{\alpha^i} Z_{A^i} \,
\bar{\alpha} \,. 
\label{Zdef}
\end{eqnarray}
Notationally we include a superscript $i$ on the diagonal fields in the various
labels on a renormalization constant and understand that there is no summation 
over this index. For the most part the relation of the bare to renormalized
quantity takes its canonical form. However, in order to ensure that the
renormalization is consistent with the Slavnov-Taylor identities derived from
the BRST symmetry in the MAG, the renormalization of the diagonal ghost, $c^i$,
and its anti-ghost is different as indicated, \cite{15}. The upshot is that one 
cannot deduce $Z_{c^i}$ from the diagonal ghost $2$-point function. To 
understand this the renormalization constant associated with the off-diagonal
ghost $2$-point function is given by the product of the renormalization
constants deriving from the external fields. From (\ref{Zdef}) this is
clearly $Z_c$. However, for the diagonal ghost the analogous product of the
wave function renormalization constants for the diagonal ghost and 
anti-ghost is unity, \cite{15}. In other words the diagonal ghost $2$-point
function is finite and $Z_{c^i}$ can only be deduced from another Green's
function which has to be a $3$-point function. Moreover, it has to be a vertex
which has strictly only one diagonal ghost or anti-ghost. In \cite{47} we used 
the $A^a_\mu {\bar c}^i c^b$ vertex for this renormalization. As the vertices
are ordinarily used to extract the coupling constant renormalization this means
that to determine the $l$th loop anomalous dimension for $c^i$ one has to 
renormalize the $A^a_\mu {\bar c}^i c^b$ vertex at the $(l+1)$th order, 
\cite{15,47}. This is on the assumption that the coupling constant 
renormalization constant has already been set in a particular scheme. The other
feature from the algebraic renormalization analysis is that $Z_{A^i}$ is in 
effect the same as the coupling constant renormalization, \cite{15}. As was 
noted in \cite{15,47} this suggests a particular similarity with the background
field gauge developed in \cite{41,42,43,44,45,46} where the $\beta$-function is
given by the background gluon wave function renormalization. More importantly 
for our MOM analysis the focus of our computations will be on computing the 
renormalization constants for the off-diagonal fields, and thence the coupling 
constant renormalization for the vertices defining a MOM scheme, as these are 
not determined from any Slavnov-Taylor identity. 

As we will be concentrating on the higher order renormalization of QCD in the
MAG we need to review the relevant properties of the renormalization group
equation. First, we recall the definition of the renormalization group 
functions for the fields, denoted generically by $\phi$, and $\alpha$ are  
\begin{equation}
\gamma_\phi(a,\alpha) ~=~ \mu \frac{\partial~}{\partial\mu} \ln Z_\phi ~~,~~
\gamma_\alpha(a,\alpha) ~=~ \frac{\mu}{\alpha} 
\frac{\partial \alpha}{\partial \mu} ~.
\end{equation}
With 
\begin{equation}
\mu \frac{\partial~}{\partial\mu} ~=~ \beta(a,\alpha) 
\frac{\partial~}{\partial a} ~+~ \alpha \gamma_\alpha(a,\alpha) 
\frac{\partial~}{\partial \alpha} 
\label{rgedef}
\end{equation}
we have 
\begin{eqnarray}
\gamma_A(a,\alpha) &=& \beta(a,\alpha) \frac{\partial}{\partial a} \ln Z_A ~+~ 
\alpha \gamma_\alpha(a,\alpha) \frac{\partial}{\partial \alpha} \ln Z_A 
\nonumber \\
\gamma_\alpha(a,\alpha) &=& \left[ \beta(a,\alpha) \frac{\partial}{\partial a}
\ln Z_\alpha ~-~ \gamma_A(a,\alpha) \right] \left[ 1 ~-~ \alpha
\frac{\partial}{\partial \alpha} \ln Z_\alpha \right]^{-1} \nonumber \\
\gamma_{A^i}(a,\alpha) &=& \beta(a,\alpha) \frac{\partial}{\partial a} 
\ln Z_{A^i} ~+~ \alpha \gamma_\alpha(a,\alpha) \frac{\partial}{\partial \alpha}
\ln Z_{A^i} \nonumber \\
\gamma_c(a,\alpha) &=& \beta(a,\alpha) \frac{\partial}{\partial a} \ln Z_c ~+~ 
\alpha \gamma_\alpha(a,\alpha) \frac{\partial}{\partial \alpha} \ln Z_c 
\nonumber \\
\gamma_{c^i}(a,\alpha) &=& \beta(a,\alpha) \frac{\partial}{\partial a}
\ln Z_{c^i} ~+~ \alpha \gamma_\alpha(a,\alpha) 
\frac{\partial}{\partial \alpha} \ln Z_{c^i} \nonumber \\
\gamma_\psi(a,\alpha) &=& \beta(a,\alpha) \frac{\partial}{\partial a} 
\ln Z_\psi ~+~ \alpha \gamma_\alpha(a,\alpha) \frac{\partial}{\partial \alpha} 
\ln Z_\psi 
\label{rgedefcon}
\end{eqnarray}
where $a$~$=$~$g^2/(16\pi^2)$. Some clarification is perhaps in order for the
forms of $\gamma_A(a,\alpha)$ and $\gamma_\alpha(a,\alpha)$. If one was
working in a linear covariant gauge such as the Landau gauge the gauge 
parameter does not get renormalized and $Z_\alpha$~$=$~$1$ in our conventions.
Therefore the second equation of (\ref{rgedefcon}) would reflect the textbook
situation if one formally sets $Z_\alpha$~$=$~$1$. In nonlinear covariant
gauges such as the Curci-Ferrari gauge and the MAG $Z_\alpha$~$\neq$~$1$. 
Therefore one has to be careful in deriving (\ref{rgedefcon}) from 
(\ref{rgedef}) in order to express $\gamma_A(a,\alpha)$ and 
$\gamma_\alpha(a,\alpha)$ purely in terms of their respective renormalization
constants $Z_A$ and $Z_\alpha$. In (\ref{rgedefcon}) we have included gauge 
parameter dependence in the $\beta$-function, $\beta(a,\alpha)$, because in 
mass dependent renormalization schemes, such as the MOM ones, the 
$\beta$-function is not independent of the gauge parameter. In mass independent
schemes such as $\MSbar$ the $\beta$-function is independent of $\alpha$, 
\cite{54}. Also the definition of the renormalization group function for 
$\alpha$ is more involved than in a linear covariant gauge because by contrast 
in the MAG and Curci-Ferrari gauge $Z_\alpha$ is not equivalent to $Z_A$.

In providing (\ref{rgedefcon}) we note that these are valid in any 
renormalization scheme. However, the parameters which the renormalization group
functions depend on are defined with respect to a scheme which here will either
be a MOM scheme or the $\MSbar$ scheme. As one of our aims is to establish the 
three loop MOM renormalization group functions we must record the relation 
between parameters in different schemes and then the way of deriving the three 
loop MOM results from the two loop vertex function renormalization. For the 
first part of this exercise the relation between the coupling constant and 
gauge parameter in two schemes are given by
\begin{equation}
g_{\MOMis}(\mu) ~=~ \frac{Z_g^{\MSbars}}{Z_g^{\MOMis}} \, 
g_{\MSbars}(\mu) ~~~,~~~
\alpha_{\MOMis}(\mu) ~=~ \frac{Z_A^{\MSbars}Z_\alpha^{\MOMis}}
{Z_A^{\MOMis}Z_\alpha^{\MSbars}} \, \alpha_{\MSbars}(\mu) 
\label{paramdef}
\end{equation}
where $\MOMi$ indicates one of the MOM schemes. In practical terms one has to
be careful in deriving the relationship between the parameters since the 
renormalization constants are functions of the parameters in the scheme defined
by the label. Therefore, one constructs the perturbative relation order by 
order in the coupling constant expansion to ensure that there are no 
singularities in the regularizing parameter. Throughout we dimensionally
regularize the theory in $d$~$=$~$4$~$-$~$2\epsilon$ dimensions where 
$\epsilon$ is the regulator. Once these mappings of the parameters between the
schemes have been determined we can define the conversion functions  
$C^{\MOMis}_\phi(a,\alpha)$, where $\phi$ indicates the appropriate field, and
$C^{\MOMis}_\alpha(a,\alpha)$. These are at the core of the three loop MOM 
renormalization group construction and are defined by
\begin{equation}
C^{\MOMis}_\phi(a,\alpha) ~=~ \frac{Z_\phi^{\MOMis}}{Z_\phi^{\MSbars}}
\label{conphi}
\end{equation}
for the fields and
\begin{equation}
C^{\MOMis}_\alpha(a,\alpha) ~=~ \frac{Z_\alpha^{\MOMis}Z_A^{\MSbars}}
{Z_\alpha^{\MSbars}Z_A^{\MOMis}}
\label{conal}
\end{equation}
for the gauge parameter. As has been our convention, \cite{37}, the variables
$a$ and $\alpha$ are $\MSbar$ parameters. In (\ref{conphi}) and (\ref{conal})
the coupling constant and gauge parameter dependence has been omitted for 
reasons of space. In each the dependence is given by 
\begin{eqnarray}
Z_\phi^{\MOMis} &=& Z_\phi^{\MOMis}
\left( a_{\mbox{$\MOMis$}}(a,\alpha) , \alpha_{\mbox{$\MOMis$}} (a,\alpha)
\right) \nonumber \\
Z_\alpha^{\MOMis} &=& Z_\alpha^{\MOMis}
\left( a_{\mbox{$\MOMis$}}(a,\alpha) , \alpha_{\mbox{$\MOMis$}} (a,\alpha)
\right) 
\end{eqnarray}
because we have chosen the $\MSbar$ scheme as the reference scheme. In
computing the explicit forms for the conversion functions from the 
renormalization constants at a particular order one has to use the relation 
between each of the parameters which was determined at the previous order. This
iterative procedure then ensures that the conversion functions are finite with
respect to $\epsilon$. With these the formal relation of the renormalization 
group functions in different schemes is 
\begin{eqnarray}
\beta^{\mbox{$\MOMis$}} ( a_{\mbox{$\MOMis$}}, \alpha_{\mbox{$\MOMis$}} ) &=&
\left[ \beta^{\mbox{$\MSbars$}}( a_{\mbox{$\MSbars$}} )
\frac{\partial a_{\mbox{$\MOMis$}}}{\partial a_{\mbox{$\MSbars$}}} \,+\,
\alpha_{\mbox{$\MSbars$}} \gamma^{\mbox{$\MSbars$}}_\alpha
( a_{\mbox{$\MSbars$}}, \alpha_{\mbox{\footnotesize{$\MSbars$}}} )
\frac{\partial a_{\mbox{$\MOMis$}}}{\partial \alpha_{\mbox{$\MSbars$}}}
\right]_{ \MSbars \rightarrow \MOMis } \nonumber \\
\gamma_\phi^{\MOMis} ( a_{\MOMis}, \alpha_{\MOMis} )
&=& \!\! \! \left[ \gamma_\phi^{\MSbars} \left(a_{\MSbars}\right)
+ \beta^{\MSbars}\left(a_{\MSbars}\right)
\frac{\partial ~}{\partial a_{\MSbars}} \ln C_\phi^{\MOMis}
\left(a_{\MSbars},\alpha_{\MSbars}\right) \right. \nonumber \\
&& \left. +~ \alpha_{\MSbars} \gamma^{\MSbars}_\alpha
\left(a_{\MSbars},\alpha_{\MSbars}\right)
\frac{\partial ~}{\partial \alpha_{\MSbars}}
\ln C_\phi^{\MOMis} \left(a_{\MSbars},\alpha_{\MSbars}\right)
\right]_{ \MSbars \rightarrow \MOMis } \nonumber \\
\end{eqnarray}
where the subscript mapping on the parentheses indicates that after the object 
is computed in $\MSbar$ variables, they are mapped to $\MOMi$ ones, \cite{55}.

\sect{Computational setup.}

Having outlined the relevant aspects of the renormalization group we now turn
to the practical aspects of the calculation. As in the previous computation, 
\cite{37}, we focus on the three vertices at the symmetric point which will 
define the three MOM schemes. They are given by  
\begin{eqnarray}
\left. \left\langle A^a_\mu(p) A^b_\nu(q) A^c_\sigma(r)
\right\rangle \right|_{p^2 = q^2 = - \mu^2} &=& f^{abc}
\left. \Sigma^{\mbox{\footnotesize{ggg}}}_{\mu \nu \sigma}(p,q)
\right|_{p^2 = q^2 = - \mu^2} \nonumber \\
\left. \left\langle c^a(p) \bar{c}^b(q) A^c_\sigma(r)
\right\rangle \right|_{p^2 = q^2 = - \mu^2} &=& f^{abc}
\left. \Sigma^{\mbox{\footnotesize{ccg}}}_\sigma(p,q)
\right|_{p^2 = q^2 = - \mu^2} \nonumber \\
\left. \left\langle \psi(p) \bar{\psi}(q) A^c_\sigma(r)
\right\rangle \right|_{p^2 = q^2 = - \mu^2} &=& T^c
\left. \Sigma^{\mbox{\footnotesize{qqg}}}_\sigma(p,q)
\right|_{p^2 = q^2 = - \mu^2}
\label{vertdecomp}
\end{eqnarray}
where $p$, $q$ and $r$ are external momenta and we choose the third momentum to
be the dependent one  
\begin{equation}
r ~=~ -~ p ~-~ q 
\end{equation}
with 
\begin{equation}
p^2 ~=~ q^2 ~=~ r^2 ~=~ -~ \mu^2
\end{equation}
defining the symmetric point giving 
\begin{equation}
pq ~=~ \frac{1}{2} \mu^2 ~.
\end{equation}
The colour group tensors for each vertex have been factored off from the
Lorentz structure $\left. \Sigma^V_{\mu_1 \ldots \mu_n}(p,q)
\right|_{p^2 = q^2 = - \mu^2}$. We note that (\ref{vertdecomp}) will be used
for the calculations in both gauges. For the Curci-Ferrari case as there are no
diagonal indices the global index $A$ used in (\ref{cffix}) can be
unambiguously identified with the index $a$. The absence of the totally 
symmetric tensor $d^{abc}$ at least in our two loop decomposition derives from
Furry's theorem and its consequences in massless QCD. Here $V$ indicates the 
appropriate vertex and $n$ is unity for the quark and ghost vertices but $3$ 
for the triple off-diagonal gluon vertex. The restriction to the symmetric 
point is included as the Lorentz structure of the full vertex away from this 
point is different. Both have been discussed in previous work, \cite{37,56}. 
For the MAG case the explicit forms of the tensors into which each vertex is 
decomposed is given in \cite{37} and we will use the same basis here for 
consistency. More specifically the Lorentz amplitude for each vertex is 
decomposed into the full basis of tensors, where the coefficients of each 
Lorentz tensor corresponds to the scalar Feynman integrals within the Green's 
functions, as 
\begin{eqnarray}
\left. \frac{}{} \Sigma^{\mbox{\footnotesize{ggg}}}_{\mu \nu \sigma}
(p,q) \right|_{p^2 = q^2 = - \mu^2} &=& \sum_{k=1}^{14}
{\cal P}^{\mbox{\footnotesize{ggg}}}_{(k) \, \mu \nu \sigma }(p,q) \,
\left( \left. \Sigma^{\mbox{\footnotesize{ggg}}}_{(k)}(p,q)
\right|_{p^2 = q^2 = - \mu^2} \right) \nonumber \\
\left. \frac{}{} \Sigma^{\mbox{\footnotesize{ccg}}}_\sigma(p,q)
\right|_{p^2 = q^2 = - \mu^2} &=& \sum_{k=1}^{2}
{\cal P}^{\mbox{\footnotesize{ccg}}}_{(k) \, \sigma }(p,q) \,
\left( \left. \Sigma^{\mbox{\footnotesize{ccg}}}_{(k)}(p,q)
\right|_{p^2 = q^2 = - \mu^2} \right) \nonumber \\
\left. \frac{}{} \Sigma^{\mbox{\footnotesize{qqg}}}_\sigma(p,q)
\right|_{p^2 = q^2 = - \mu^2} &=& \sum_{k=1}^{6}
{\cal P}^{\mbox{\footnotesize{qqg}}}_{(k) \, \sigma }(p,q) \,
\left( \left. \Sigma^{\mbox{\footnotesize{qqg}}}_{(k)}(p,q) 
\right|_{p^2 = q^2 = - \mu^2} \right)
\end{eqnarray}
where $k$ labels a particular tensor. To extract the perturbative expansion for
the scalar amplitudes we use the projection method which was discussed in 
\cite{37}. Briefly to determine a particular amplitude one multiplies each 
vertex function by a specific linear combination of tensors from the basis,
\begin{eqnarray}
f^{abc} \Sigma^{\mbox{\footnotesize{ggg}}}_{(k)}(p,q)
&=& {\cal M}^{\mbox{\footnotesize{ggg}}}_{kl} \left(
{\cal P}^{\mbox{\footnotesize{ggg}} \, \mu \nu \sigma}_{(l)}(p,q) \left.
\left\langle A^a_\mu(p) A^b_\nu(q) A^c_\sigma(r)
\right\rangle \right)\right|_{p^2 = q^2 = - \mu^2} \nonumber \\
f^{abc} \Sigma^{\mbox{\footnotesize{ccg}}}_{(k)}(p,q) &=&
{\cal M}^{\mbox{\footnotesize{ccg}}}_{kl} \left(
{\cal P}^{\mbox{\footnotesize{ccg}} \, \sigma}_{(l)}(p,q) \left.
\left\langle c^a(p) \bar{c}^b(q) A^c_\sigma(r)
\right\rangle \right) \right|_{p^2 = q^2 = - \mu^2} \nonumber \\
T^c \Sigma^{\mbox{\footnotesize{qqg}}}_{(k)}(p,q) &=&
{\cal M}^{\mbox{\footnotesize{qqg}}}_{kl} \left(
{\cal P}^{\mbox{\footnotesize{qqg}} \, \sigma}_{(l)}(p,q) \left.
\left\langle \psi(p) \bar{\psi}(q) A^c_\sigma(r)
\right\rangle \right) \right|_{p^2 = q^2 = - \mu^2}
\end{eqnarray}
where ${\cal M}^V_{kl}$ is a matrix whose elements are rational polynomials in
$d$ and whose $k$th row is the linear combination required for the $k$th
amplitude. This matrix is given in \cite{37} for each vertex. The colour group
dependence has been included here to balance the colour indices on the right
hand side. As noted earlier to two loops the left hand side reflects the
actual structure. If it were not the case then we would have to introduce a
colour projection. In performing the Lorentz projection the Lorentz integrals 
within each vertex function become scalar integrals and the resulting numerator
scalar products are rewritten as far as possible in terms of the propagators. 
The reason for this is that we will use the Laporta algorithm, \cite{57}, to 
perform the computations. This is a method which derives integration by parts 
relations between scalar Feynman integrals and then solves them in terms of a 
relatively small set of master integrals. The values of these masters are 
determined by direct methods. In rewriting the scalar products as indicated 
this may produce an irreducible numerator. One feature of the Laporta 
algorithm, \cite{57}, is that it can handle such irreducible cases 
systematically. For our specific $3$-point symmetric vertex computation there 
is one topology in the one loop integral family which is the triangle graph. At
two loops there are two topologies in that integral family. One is the two loop
non-planar vertex and the other is the ladder graph. These are illustrated in 
Figure $1$. 
{\begin{figure}
\begin{center}
\SetScale{1.2}
\begin{picture}(240,100)(10,-10)
\Line(-20,0)(10,60)
\Line(40,0)(10,60)
\Line(-15,10)(35,10)
\Line(10,60)(10,75)
\Line(80,0)(110,60)
\Line(140,0)(110,60)
\Line(85,10)(105,23.3)
\Line(115,30)(122.5,35)
\Line(97.5,35)(135,10)
\Line(110,60)(110,75)
\Line(180,0)(210,60)
\Line(240,0)(210,60)
\Line(185,10)(235,10)
\Line(197.5,35)(222.5,35)
\Line(210,60)(210,75)
\end{picture}
\caption{Integral families at one and two loops for the symmetric point.}
\end{center}
\end{figure}
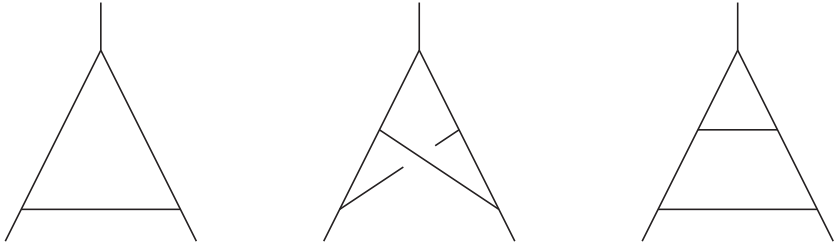}
If one was away from the symmetric point then there would be at most two 
additional ladder topologies, \cite{56}, which correspond to two rotations of
the final graph.

{\begin{table}[hb]
\begin{center}
\begin{tabular}{|c||c|c|c|c|}
\hline
Green's function & One loop & Two loop & Total \\
\hline
$A^a_\mu \, A^b_\nu$ & $~\,6$ & $~131$ & $~~137$ \\
$c^a \bar{c}^b$ & $~\,3$ & $~~\,54$ & $~~~57$ \\
$\psi \bar{\psi}$ & $~\,3$ & $~~\,81$ & $~~~84$ \\
$ A^a_\mu \, A^b_\nu A^c_\sigma$ & $23$ & $1291$ & $1314$ \\
$c^a \bar{c}^b A^c_\sigma$ & $16$ & $~867$ & $~~883$ \\
$\psi \bar{\psi} A^c_\sigma$ & $~\,5$ & $~217$ & $~~222$ \\
\hline
Total & $56$ & $2641$ & $2697$ \\
\hline
\end{tabular}
\end{center}
\begin{center}
{Table 1. Number of Feynman diagrams for each $2$- and $3$-point function in
the MAG.}
\end{center}
\end{table}}

In terms of practicalities such a computation can only be managed within a 
reasonable amount of time with the use of computer algebra packages. The main
tool for handling the large amounts of tedious algebra is {\sc Form}, 
\cite{58}, and its threaded version {\sc Tform}, \cite{59}. The Feynman graphs 
are generated using the {\sc Qgraf} package \cite{60} and then converted into
{\sc Form} notation where all the colour and Lorentz indices are added. The 
number of graphs computed for each vertex is given in Table $1$ for the MAG and
Table $2$ for the Curci-Ferrari gauge. For the implementation of the Laporta 
algorithm we have chosen to use the {\sc Reduze} package, \cite{61}, which is 
written in {\sc GiNaC}, \cite{62}. One useful feature of {\sc Reduze} is that 
the reduction to master integrals can be extracted from the database of 
relations {\sc Reduze} creates in {\sc Form} syntax. This has allowed us to set
up an automatic computation whereby the relevant integrals from the database 
are included within a {\sc Form} module. The remaining general tasks are the 
evaluation and inclusion of the master integrals and the renormalization. For 
the former all the one and two loop masters are already known to the order in 
$\epsilon$ required for the two loop vertex functions to the finite part, 
\cite{63,64,65,66,67,68}. A complete set for easy reference has been provided 
in \cite{67} and we use the same notation throughout this article. For several 
masters the expansion in $\epsilon$ is needed to $O(\epsilon^2)$. Ordinarily 
for a two loop renormalization this would not be necessary. However, in the 
construction of the integration by parts relations spurious poles in $\epsilon$
appear which multiply several masters. This requires the extra terms in the 
master integral $\epsilon$ expansion. While we will be discussing general 
features of the full analytic results later with the explicit expressions being
included in attached data files we need to comment on the structure. This is 
dictated by the expressions for the masters and involve the polylogarithm 
function $\mbox{Li}_n(z)$ via the function  
\begin{equation}
s_n(z) ~=~ \frac{1}{\sqrt{3}} \Im \left[ \mbox{Li}_n \left(
\frac{e^{iz}}{\sqrt{3}} \right) \right] ~.
\end{equation}
In previous work in other gauges, \cite{69}, the final expressions involved the
quantity $\Sigma$ which was defined as the following combination of harmonic 
polylogarithms
\begin{equation}
\Sigma ~=~ {\cal H}^{(2)}_{31} ~+~ {\cal H}^{(2)}_{43} 
\end{equation}
in the notation of \cite{67}. Such quantities are not unrelated to harmonic
polylogarithms based on cyclotomic polynomials, \cite{70}. However, it 
transpires that this object was not independent of another combination of 
quantities which appear since, \cite{68}, 
\begin{equation}
\Sigma ~=~ \frac{1}{36} \psi^{\prime\prime\prime}\left( \frac{1}{3} \right) ~-~
\frac{2\pi^4}{27} 
\end{equation}
where $\psi(z)$ is the derivative of the logarithm of the Euler 
$\Gamma$-function. Therefore, in the expressions in our data files the object 
$\Sigma$ does not formally appear unlike \cite{69}. Of course in numerical 
results both quantities have the same value. To assist numerical evaluation we 
note that  
\begin{eqnarray}
\zeta_3 &=& 1.20205690 ~~,~~
\psi^\prime ( \third ) ~=~ 10.09559713 ~~,~~
\psi^{\prime\prime\prime} ( \third ) ~=~ 488.1838167 \nonumber \\
s_2 ( \pitwo ) &=& 0.32225882 ~~,~~
s_2 ( \pisix ) ~=~ 0.22459602 ~~,~~
s_3 ( \pitwo ) ~=~ 0.32948320 \nonumber \\
s_3 ( \pisix ) &=& 0.19259341 
\end{eqnarray}
where $\zeta_z$ is the Riemann zeta function. Finally, as we are performing an 
automatic symbolic manipulation programme we use the renormalization procedure 
developed in \cite{71} to extract the renormalization constants for each 
vertex. Briefly all vertex functions are computed in terms of bare parameters 
which means the coupling constant and gauge parameter. Their associated 
counterterms are introduced symbolically after all graphs have been computed 
and summed by rescaling with the appropriate renormalization constant defined 
in (\ref{Zdef}). 

{\begin{table}[ht]
\begin{center}
\begin{tabular}{|c||c|c|c|c|}
\hline
Green's function & One loop & Two loop & Total \\
\hline
$A^A_\mu \, A^B_\nu$ & $~3$ & $~\,19$ & $~\,22$ \\
$c^A \bar{c}^B$ & $~1$ & $~~~9$ & $~\,10$ \\
$\psi \bar{\psi}$ & $~1$ & $~~~6$ & $~~~7$ \\
$ A^A_\mu \, A^B_\nu A^C_\sigma$ & $~8$ & $112$ & $120$ \\
$c^A \bar{c}^B A^C_\sigma$ & $~3$ & $~\,49$ & $~\,52$ \\
$\psi \bar{\psi} A^C_\sigma$ & $~2$ & $~\,33$ & $~\,35$ \\
\hline
Total & $18$ & $228$ & $246$ \\
\hline
\end{tabular}
\end{center}
\begin{center}
{Table 2. Number of Feynman diagrams for each $2$- and $3$-point function in
the Curci-Ferrari gauge.}
\end{center}
\end{table}}

To extract the MOM renormalization constant from each vertex function
additionally requires the wave function renormalization constant of the 
external fields in the MOM scheme. This is achieved by performing the $2$-point
function two loop renormalization of the off-diagonal gluon, ghost and quark 
fields in each of the MOM schemes. For these we use the {\sc Mincer} algorithm,
\cite{72}, which is implemented in {\sc Form}, \cite{73}. The number of graphs 
for each of the $2$-point functions is given in Table $1$ for the MAG and those
for the Curci-Ferrari gauge are given in Table $2$. In extracting the 
wave function and gauge parameter renormalization constants, using the same 
automatic procedure as \cite{71}, we note that the one loop $2$-point functions
are renormalized first in the MOM scheme of \cite{38,39} and then the one loop 
vertex functions. The latter define the three schemes which are then used to
determine the wave function and gauge parameter renormalizations at two loops
before these are used to deduce the coupling constant renormalization constants
for each of the three MOM schemes. We note that the method to define each MOM 
scheme is based on the original programme of \cite{38,39} and was followed in 
\cite{37}. For each of the $2$-point and vertex function renormalizations at
the subtraction point the MOM scheme is defined so that after the 
renormalization constant has been defined there are no $O(a)$ corrections. For 
the vertex functions this is qualified by noting that it is the Lorentz 
channels of the tree level which has no $O(a)$ corrections after 
renormalization. The non-tree level vertex structures will have $O(a)$ 
corrections at the symmetric point. As one check on our computer algebraic 
programmes we have verified that the two loop $\MSbar$ coupling constant 
renormalization constant of \cite{74,75} correctly emerges from each $3$-point 
vertex function. This completes the description of the technology to compute 
the $3$-point functions at the symmetric point. It now remains to discuss the 
results. 

\sect{Results.}

Before discussing the renormalization group functions and vertex functions we
detail the additional checks on our computations. As the first stage in 
considering the renormalization of the MAG and Curci-Ferrari gauges beyond that
of \cite{37} in MOM schemes, we have determined each vertex function in the 
$\MSbar$ scheme at the symmetric point. An important check on the computations 
is that at the symmetric point the divergent terms in $\epsilon$ can be 
minimally subtracted and the resulting renormalization constants agree with 
those of \cite{47,76}. By this we mean that the wave function renormalization 
constants associated with the external legs of the respective vertex functions 
are such that the final renormalization constant corresponding to the coupling 
constant correctly emerges in agreement with the known two loop $\MSbar$ result
of \cite{1,2,74,75}. An additional check is that the relations between various 
amplitudes which were observed in \cite{37} at one loop are maintained at two 
loops. For instance, those of the triple off-diagonal gluon in the MAG are the 
same as those of the triple gluon in the linear covariant gauge. Thus at the 
symmetric point we have checked that the relations 
\begin{eqnarray}
\left. \Sigma^{\mbox{\footnotesize{ggg}}}_{(1)}(p,q) 
\right|_{p^2=q^2=-\mu^2}^{\MSbars} &=&
\left. \Sigma^{\mbox{\footnotesize{ggg}}}_{(2)}(p,q) 
\right|_{p^2=q^2=-\mu^2}^{\MSbars} ~=~
-~ \frac{1}{2} \left. \Sigma^{\mbox{\footnotesize{ggg}}}_{(3)}(p,q)
\right|_{p^2=q^2=-\mu^2}^{\MSbars} \nonumber \\
&=& -~ \left. \Sigma^{\mbox{\footnotesize{ggg}}}_{(4)}(p,q)
\right|_{p^2=q^2=-\mu^2}^{\MSbars} ~=~ 
\frac{1}{2} \left. \Sigma^{\mbox{\footnotesize{ggg}}}_{(5)}(p,q)
\right|_{p^2=q^2=-\mu^2}^{\MSbars} \nonumber \\
&=& -~ \left. \Sigma^{\mbox{\footnotesize{ggg}}}_{(6)}(p,q) 
\right|_{p^2=q^2=-\mu^2}^{\MSbars}
\nonumber \\
\left. \Sigma^{\mbox{\footnotesize{ggg}}}_{(7)}(p,q) 
\right|_{p^2=q^2=-\mu^2}^{\MSbars} &=&
2 \left. \Sigma^{\mbox{\footnotesize{ggg}}}_{(9)}(p,q) 
\right|_{p^2=q^2=-\mu^2}^{\MSbars} ~=~
-~ 2 \left. \Sigma^{\mbox{\footnotesize{ggg}}}_{(11)}(p,q)
\right|_{p^2=q^2=-\mu^2}^{\MSbars} \nonumber \\
&=& -~ \left. \Sigma^{\mbox{\footnotesize{ggg}}}_{(14)}(p,q) 
\right|_{p^2=q^2=-\mu^2}^{\MSbars}
\nonumber \\
\left. \Sigma^{\mbox{\footnotesize{ggg}}}_{(8)}(p,q) 
\right|_{p^2=q^2=-\mu^2}^{\MSbars} &=&
-~ \left. \Sigma^{\mbox{\footnotesize{ggg}}}_{(13)}(p,q) 
\right|_{p^2=q^2=-\mu^2}^{\MSbars} \nonumber \\
\left. \Sigma^{\mbox{\footnotesize{ggg}}}_{(10)}(p,q) 
\right|_{p^2=q^2=-\mu^2}^{\MSbars} &=&
-~ \left. \Sigma^{\mbox{\footnotesize{ggg}}}_{(12)}(p,q) 
\right|_{p^2=q^2=-\mu^2}^{\MSbars}
\label{msggg}
\end{eqnarray}
emerge correctly to two loops for the triple off-diagonal gluon vertex. For the 
off-diagonal ghost vertex there are two amplitudes but the nature of the vertex
in the MAG is such that only one is independent. Therefore, we found 
\begin{equation}
\left. \Sigma^{\mbox{\footnotesize{ccg}}}_{(1)}(p,q) 
\right|_{p^2=q^2=-\mu^2}^{\MSbars} ~=~
-~ \left. \Sigma^{\mbox{\footnotesize{ccg}}}_{(2)}(p,q) 
\right|_{p^2=q^2=-\mu^2}^{\MSbars} ~.
\label{msccg}
\end{equation} 
Finally, for the quark off-diagonal gluon vertex we have verified that 
\begin{equation}
\left. \Sigma^{\mbox{\footnotesize{qqg}}}_{(2)}(p,q) 
\right|_{p^2=q^2=-\mu^2}^{\MSbars} ~=~
\left. \Sigma^{\mbox{\footnotesize{qqg}}}_{(5)}(p,q) 
\right|_{p^2=q^2=-\mu^2}^{\MSbars} ~\,,~\,
\left. \Sigma^{\mbox{\footnotesize{qqg}}}_{(3)}(p,q) 
\right|_{p^2=q^2=-\mu^2}^{\MSbars} ~=~
\left. \Sigma^{\mbox{\footnotesize{qqg}}}_{(4)}(p,q) 
\right|_{p^2=q^2=-\mu^2}^{\MSbars}
\label{msqqg}
\end{equation} 
are satisfied like the others for all values of $\alpha$. The amplitudes 
associated with channels $1$ and $6$ in the quark-gluon vertex are not related 
to any of the others. The former corresponds to the tree level vertex and the 
latter is in a separate partition of spinor space as discussed in \cite{69}. 
One feature which is apparent in MAG expressions, and those at one loop in 
\cite{37}, is that the amplitudes corresponding to the original Feynman rule 
are non-singular in $\alpha$. Thus using this channel for the definition of MOM
schemes does not lead to problems in the true definition of the MAG. For the 
Curci-Ferrari gauge the same relations between the amplitudes hold. For the 
ghost-gluon vertex this is different from the situation in the linear covariant
gauge. In that gauge the ghost-gluon vertex is not antisymmetric since the 
spacetime derivative in the Lagrangian only acts on one of the ghost fields 
unlike the Curci-Ferrari gauge. Thus in the latter the amplitudes are related 
as given above. 

Now that the evaluation of the vertex functions have been established in the
$\MSbar$ scheme and the correct renormalization group functions emerge we turn
to the situation in the MOM schemes. To summarize we have defined $\MOMi$ with 
respect to the Lorentz channel corresponding to the tree level vertex 
structure. In other words at the fully symmetric point the coupling constant 
renormalization constant is chosen such that there are no $O(a)$ corrections in
keeping with the ethos of \cite{38,39}. The process is an iterative one. 
Briefly at a given loop order all $2$-point functions are first rendered finite
in $\MOMi$. Then the appropriate $\MOMi$ vertex is renormalized at the same 
loop order. Once equipped with this coupling constant renormalization constant,
the subsequent loop order of all the $2$-point functions are renormalized in 
$\MOMi$ before repeating the exercise for the coupling constant 
renormalization. This establishes the $\MOMi$ renormalization constants at two
loops and then we deduce the various conversion functions to two loops. These 
are required for going beyond this order to determine the {\em three} loop 
renormalization group functions ahead of an explicit computation. In order to 
achieve this we require the mappings of the parameters between the schemes 
which are formally defined in (\ref{paramdef}). 

There are various checks on the full analytic expressions for these 
renormalization group functions. The first is that the two loop results agree
with those determined in \cite{37}. The method we used in \cite{37} was to
exploit the properties of the renormalization group. In other words the one 
loop vertex function renormalization in the MOM schemes produced the conversion
functions which, via the renormalization group formalism, determined the then 
to be explicitly computed two loop anomalous dimensions. Therefore using this 
blind check it is satisfying to record that the explicit computation is in 
agreement. The other main check is due to the relation the MAG has with the 
Curci-Ferrari gauge. If one takes the limit of the MAG where the Abelian sector
is formally removed then the remaining Lagrangian involving the off-diagonal 
fields is equivalent to the massless Curci-Ferrari Lagrangian of \cite{40}. 
Therefore, the renormalization group functions of the MAG in the 
$\Nda/\Noda$~$\rightarrow$~$0$ limit should agree with those in the 
Curci-Ferrari gauge in each of the three schemes. This is the case for 
$\MSbar$, \cite{47}. For the $\MOMi$ schemes this is also the situation here
since the three loop $\MOMi$ renormalization group functions have been 
evaluated directly in the Curci-Ferrari gauge. We note that we have taken 
the $\Nda/\Noda$~$\rightarrow$~$0$ limit in the MAG and verified that both
computations are consistent. The final check rests in the fact that the double 
poles in $\epsilon$ in the renormalization constants in the various MOM schemes are not independent and are determined by the simple pole at one loop. That the
double poles are in agreement in both gauges indicates that those graphs with 
subgraph divergences have been correctly treated within the symbolic 
manipulation programmes we have developed. Finally, the relations between the
amplitudes in (\ref{msggg}), (\ref{msccg}) and (\ref{msqqg}) for $\MSbar$ in
both gauges also hold after renormalization in the MOM schemes too. 

Having compiled all the renormalization group functions in each of the three 
schemes it is interesting to make an initial comparison of the size of the 
corrections. As a simple benchmark we consider the three loop MAG 
$\beta$-functions in each of the four schemes for $\alpha$~$=$~$0$. We have,
for instance, 
\begin{eqnarray}
\beta^{\MSbars}(a,0) &=&
\left[ 0.666667 \Nf - 11.000000 \right] a^2 ~+~ \left[ 12.666667 \Nf 
- 102.000000 \right] a^3 \nonumber \\
&& +~ \left[ - 6.018518 \Nf^2 + 279.611111 \Nf - 1428.500000 
\right] a^4 ~+~ O(a^5) \nonumber \\
\beta^{\MOMgs}(a,0) &=& 
\left[ 0.666667 \Nf - 11.000000 \right] a^2 ~+~ \left[ 12.666667 \Nf 
- 93.608510 \right] a^3 \nonumber \\
&& +~ \left[ - 2.658115 \Nf^3 + 54.791594 \Nf^2 + 401.565562 \Nf 
- 3543.358228 \right] a^4 \nonumber \\
&& +~ O(a^5) \nonumber \\
\beta^{\MOMhs}(a,0) &=&
\left[ 0.666667 \Nf - 11.000000 \right] a^2 ~+~ \left[ 12.666667 \Nf 
- 108.000000 \right] a^3 \nonumber \\
&& +~ \left[ - 25.035332 \Nf^2 + 674.085832 \Nf - 2991.050472 
\right] a^4 ~+~ O(a^5) \nonumber \\
\beta^{\MOMqs}(a,0) &=&
\left[ 0.666667 \Nf - 11.000000 \right] a^2 ~+~ \left[ 12.666667 \Nf 
- 96.936557 \right] a^3 \nonumber \\
&& +~ \left[ - 22.587812 \Nf^2 + 627.275918 \Nf - 2266.490127 
\right] a^4 ~+~ O(a^5) 
\label{betanum}
\end{eqnarray}
where the $\MSbar$ results were given originally in \cite{1,2,73,75,77}. At two
loops there is not a significant departure from the $\MSbar$ value of the 
comparable term. The major difference is in the three loop term where, for
instance, in the Yang-Mills case the coefficient in each of the three MOM 
schemes is roughly twice that of the three loop $\MSbar$ value. While for mass
independent renormalization schemes the three loop term is the first point 
where scheme dependence will arise. By contrast in mass dependent schemes, 
which includes the MOM cases, this will occur at the previous order as is 
evident in (\ref{betanum}). What is not predictable prior to an explicit 
computation is the magnitude of any correction. While the large difference with
$\MSbar$ is consistent within the three schemes, a better comparison might be 
with a physical quantity which we will consider later. From \cite{69} comparing
the same $\Nf$ independent Landau gauge coefficient in the three loop MOM 
$\beta$-functions the $\MOMg$ scheme coefficient of \cite{69} is roughly the 
same as the $\MSbar$ value but the $\MOMq$ and $\MOMh$ values are more in line 
with the analogous scheme in the MAG. This is slightly surprising as naively 
the expectation might have been that all three MOM schemes in the MAG would 
have been similar to the MOM schemes of \cite{69}. 

{\begin{figure}
\includegraphics[width=7.6cm,height=8cm]{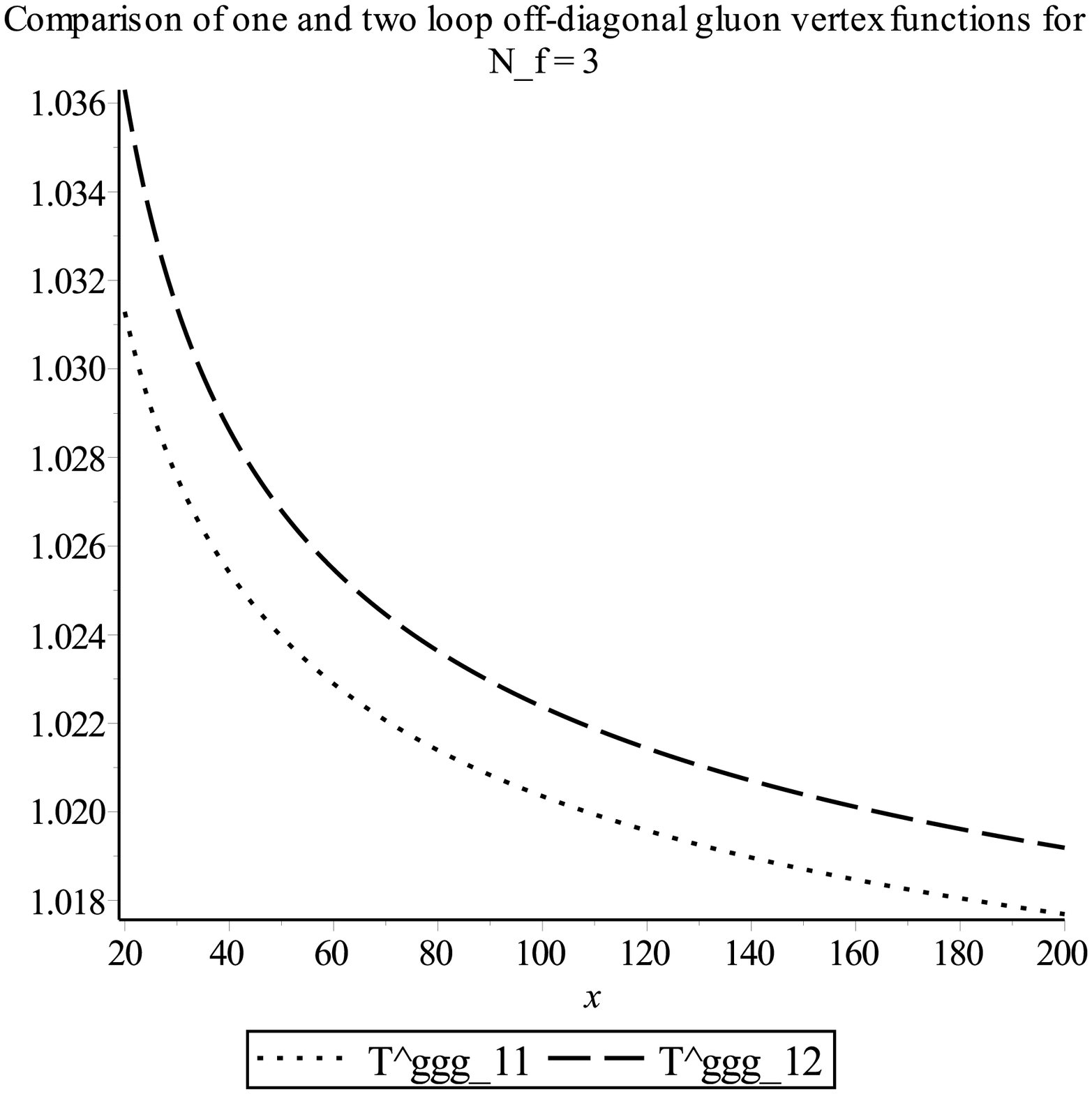}
~~~~
\includegraphics[width=7.6cm,height=8cm]{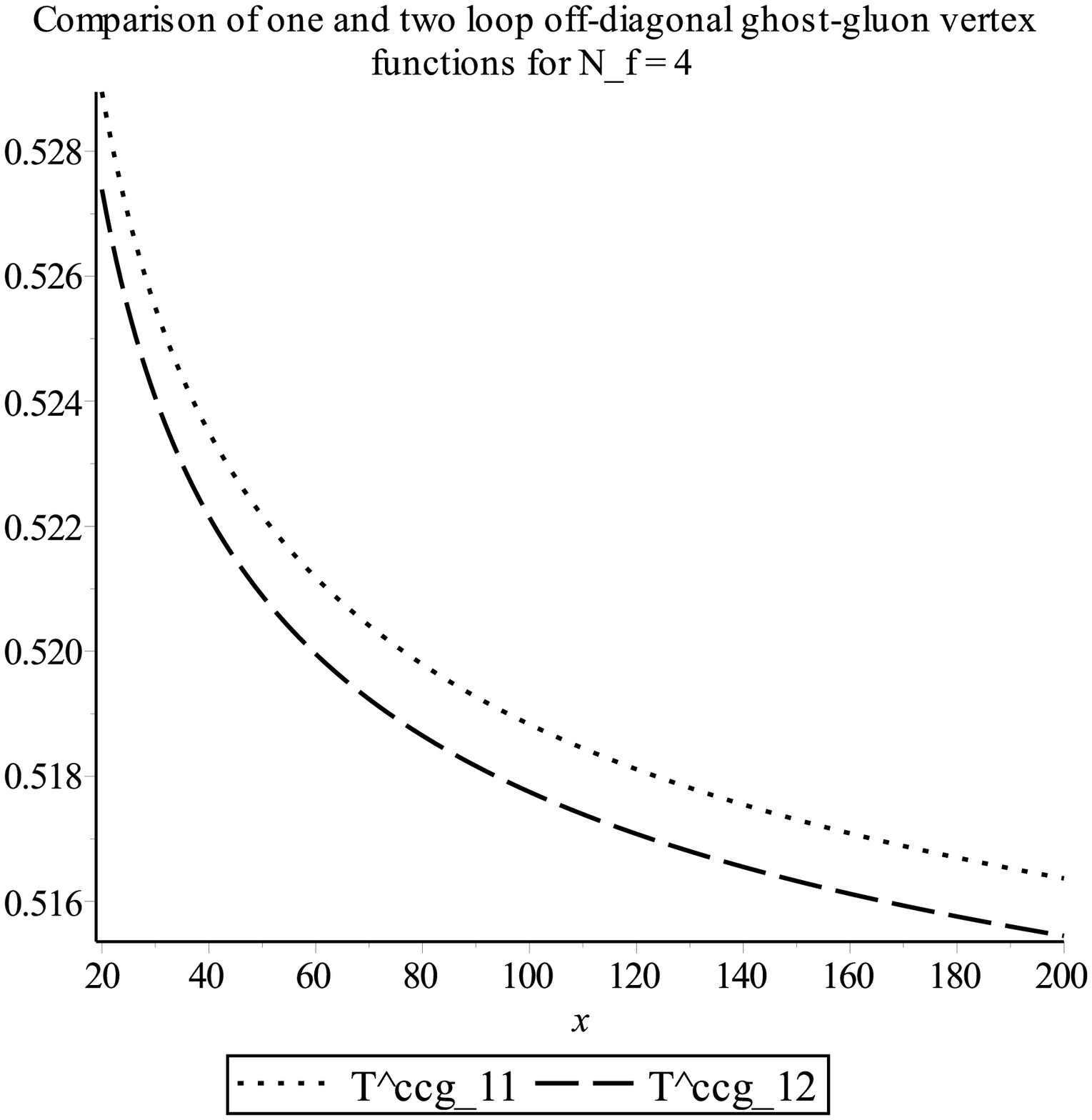}
\vspace{0.8cm}
\includegraphics[width=7.6cm,height=8cm]{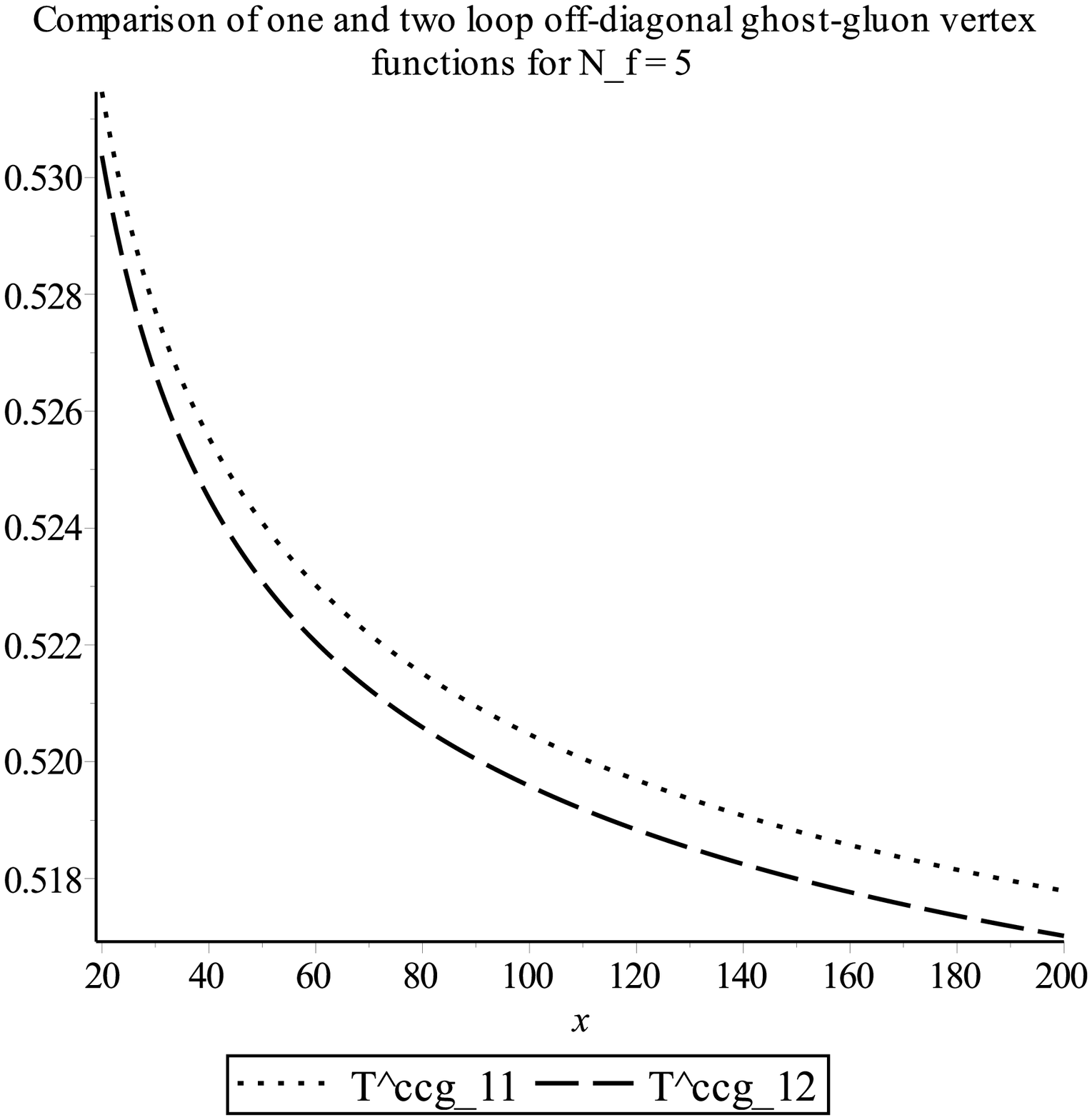}
\quad
\includegraphics[width=7.6cm,height=8cm]{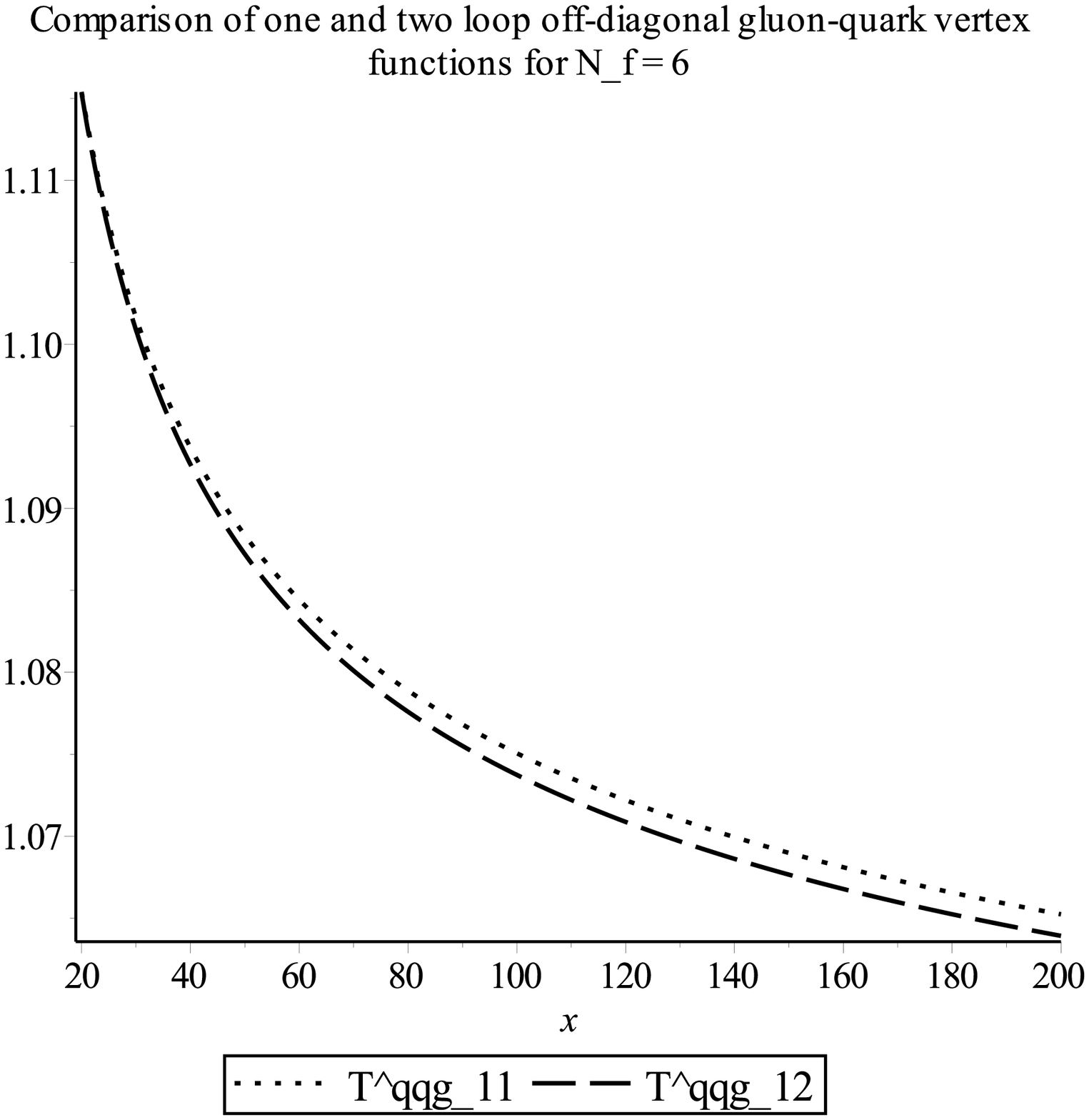}
\caption{Comparison of various $\MSbar$ MAG vertex functions for different 
values of $\Nf$.}
\end{figure}}

While the $\beta$-functions give some insight into the size of the corrections
in various schemes the effect of the higher order corrections on the structure
of the vertex functions is also of interest at the symmetric point. We have 
chosen to illustrate this graphically. So in order to construct plots of the 
vertex functions at the symmetric point with respect to a scale we first 
convert the coupling constant to its explicit scale dependence. We introduce 
the partial coupling constants $a_l(\mu,\Lambda)$, where $l$ is the loop order,
which are given by solving the $\beta$-function as a differential equation for 
the coupling constant. We have
\begin{eqnarray}
a_1(\mu,\Lambda) &=& \frac{1}{b_0 L} ~~,~~
a_2(\mu,\Lambda) ~=~ \frac{1}{b_0 L} \left[ 1 - \frac{b_1 \ln (L)}{{b_0}^2 L}
\right] \nonumber \\
a_3(\mu,\Lambda) &=& \frac{1}{b_0 L} \left[ 1 - \frac{b_1 \ln (L)}{{b_0}^2 L}
+ \left[ {b_1}^2 \left[ \ln^2 (L) - \ln (L) - 1 \right] + b_0 b_2 \right] 
\frac{1}{{b_0}^4 {L}^2} \right]
\label{partcc}
\end{eqnarray}
where
\begin{equation}
L ~=~ \ln \left( \frac{\mu^2}{{\Lambda}^2} \right) 
\end{equation}
and the $\beta$-function coefficients are defined by
\begin{equation}
\beta(a,0) ~=~ -~ \sum_{n=0}^\infty b_n a^{n+1} ~.
\end{equation}
Here $\Lambda$ is the scale associated with the constant of integration. It has
different values depending on the number of quark flavours but in this analysis
we will leave it as a free parameter and not fix it to any specific value. For 
the higher order forms of $a_l(\mu,\Lambda)$ in (\ref{partcc}) we have chosen 
to use the versions given in \cite{77} and for this analysis we will 
concentrate on the $\alpha$~$=$~$0$ case as this is the value which defines the
MAG. We will use $a_l(\mu,\Lambda)$ at the $l$th loop to construct the 
truncated vertex functions and compare them. Therefore if we write
\begin{equation}
\left. \Sigma^V_{(k)}(p,q) \right|_{p^2 = q^2 = - \mu^2} ~=~ \sum_{n=0}^\infty  
\Sigma^V_{(k)\,n} \, a^n
\end{equation}
for each vertex $V$ and channel $k$ then we define the truncated vertex 
functions $T^V_{k,l}$ at the symmetric point by
\begin{equation}
T^V_{k,l} ~=~ \sum_{n=0}^l \Sigma^V_{(k)\,n} \left( a_l(\mu,\Lambda)
\right)^n 
\end{equation}
where $l$ is the number of loops at which the truncation occurs. Having defined 
the truncated vertex functions we will give plots for $l$~$=$~$1$ and $2$ in 
the $\MSbar$ scheme at the symmetric point for the channels corresponding to 
the tree level vertex structures. This is because for the MOM schemes the 
symmetric point vertex functions are by definition a constant for all $l$ for 
the same channels. Our plots are given in Figure $2$ and we have selected a 
representative for each vertex and one of four values of $\Nf$. This is 
primarily because overall the plots are very similar in form to the ones not 
given. In general the behaviour from one to two loops is the same in that at 
higher values of $\Nf$ there is little difference between one and two loops. 
While the $\Nf$~$=$~$3$ plots suggests a larger discrepancy. Quantifying the 
difference it transpires that over the range of $x$~$=$~$\mu/\Lambda$ given in 
the Figures there is only a change of $1\%$. This is as expected as we are well
within the range of perturbative reliability. Moreover, comparing these 
corrections with comparable plots, \cite{56}, the order of the corrections is 
similar if not marginally better than those for the linear covariant gauge 
fixing. This is reassuring in light of the full off-shell two loop analysis of 
\cite{56} where it was shown that the two loop corrections were not 
significantly different from one loop for all ranges of the external momenta 
away from the symmetric point. The plots in Figure $2$ represent the diagonal 
section across the $(p^2/\mu^2,q^2/\mu^2)$ plane. In \cite{56} the vertex 
functions in a linear covariant gauge were determined over this whole plane. 

{\begin{figure}
\includegraphics[width=7.6cm,height=8cm]{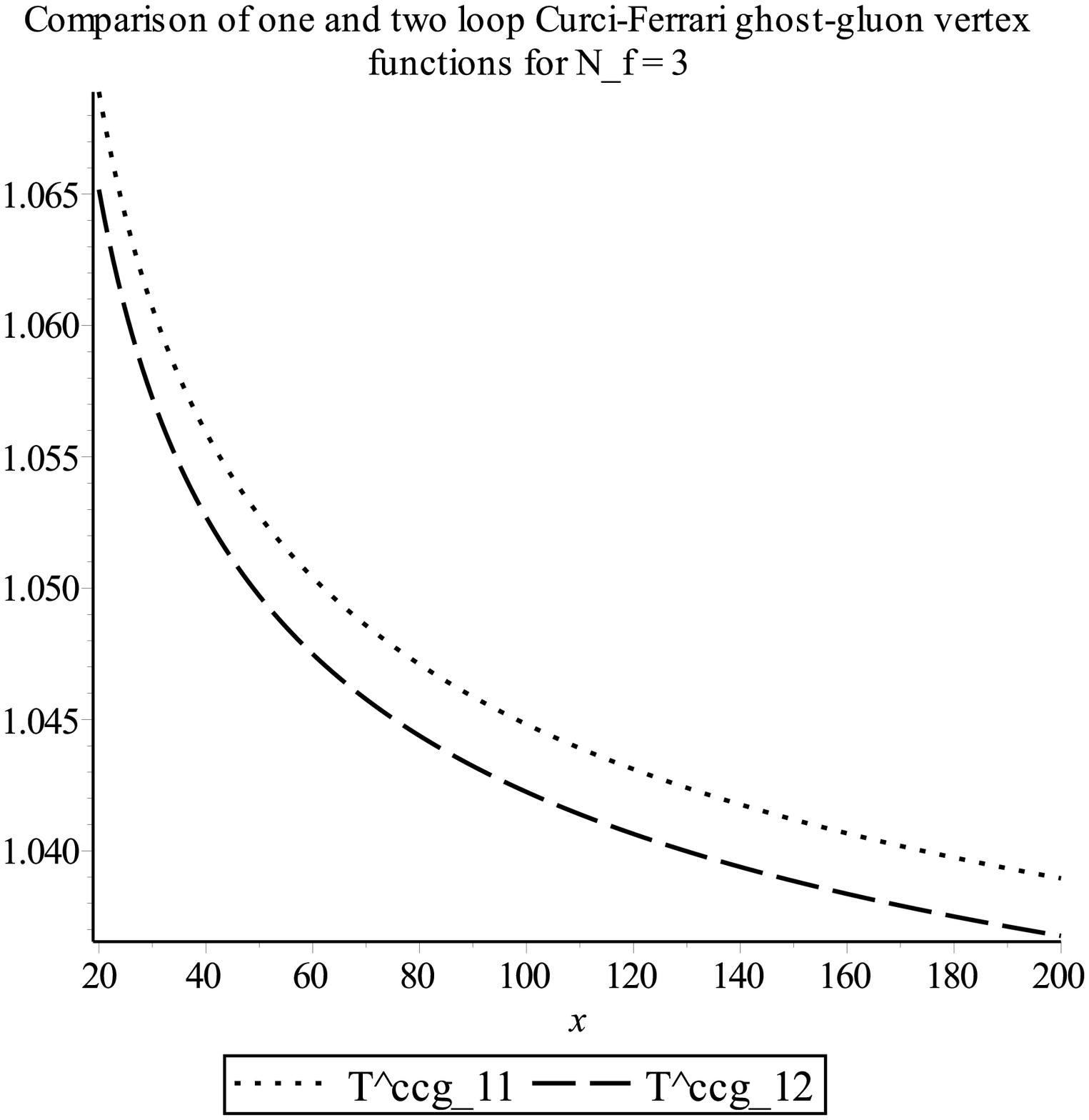}
~~~~
\includegraphics[width=7.6cm,height=8cm]{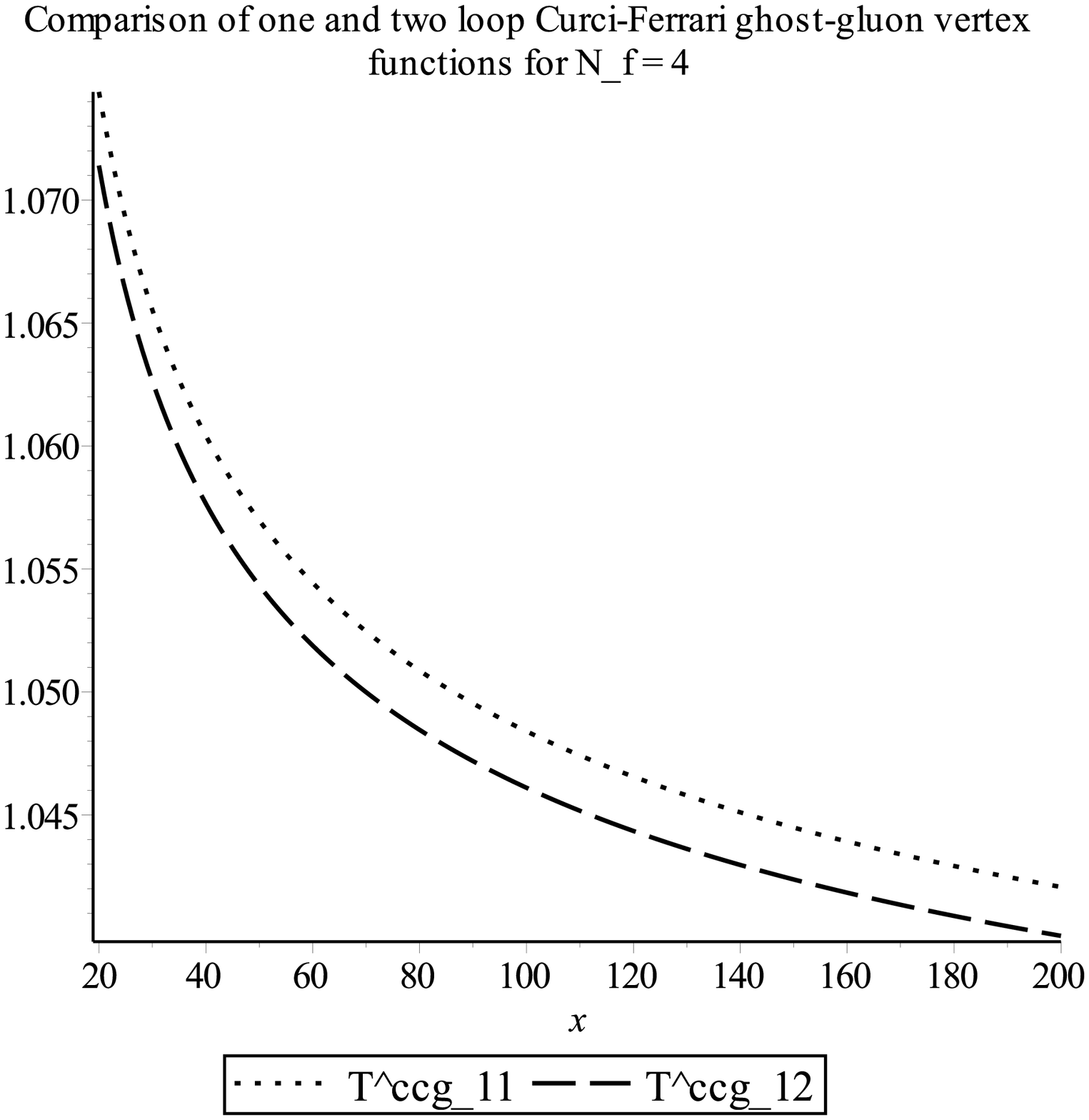}
\vspace{0.8cm}
\includegraphics[width=7.6cm,height=8cm]{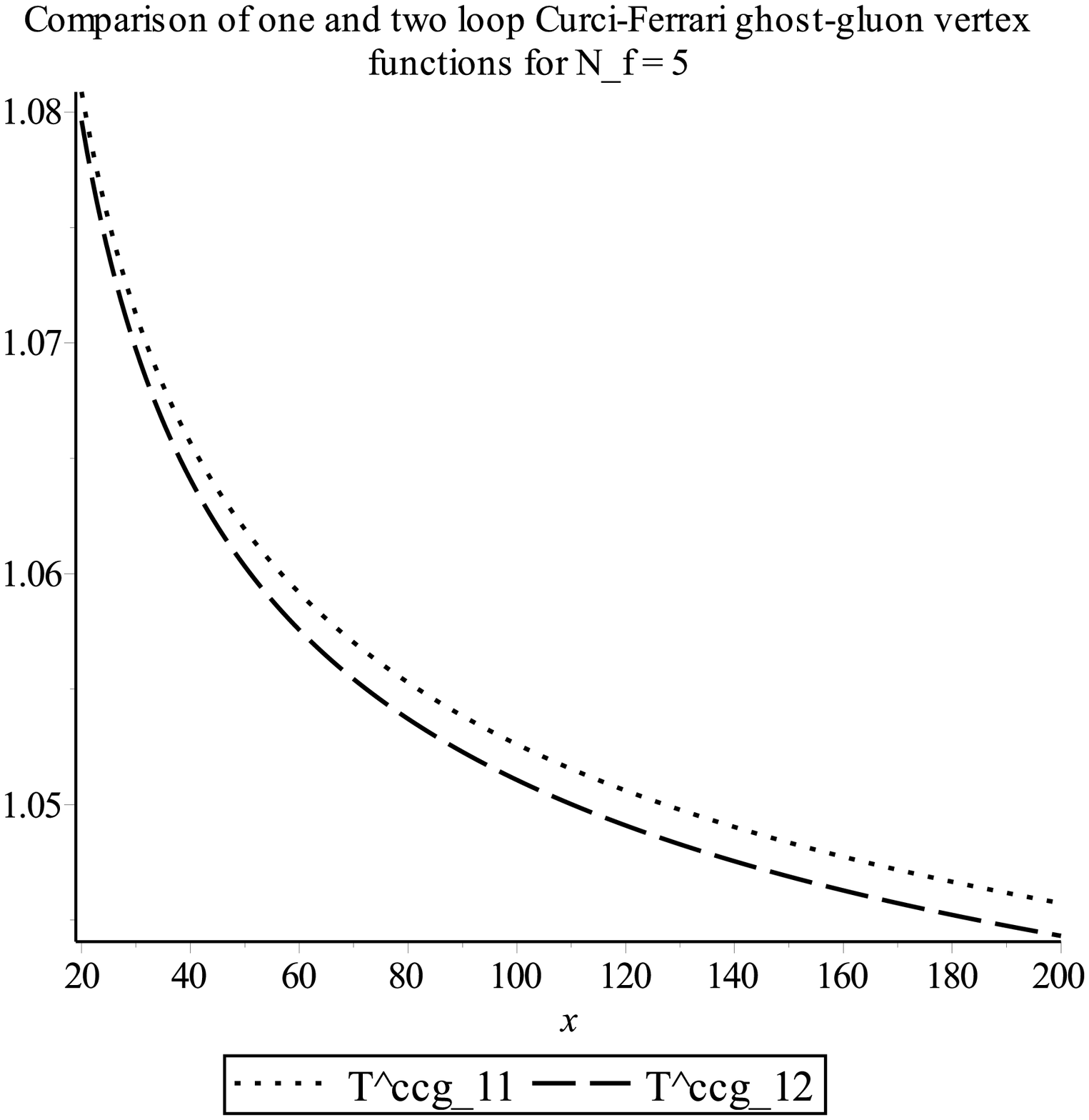}
\quad
\includegraphics[width=7.6cm,height=8cm]{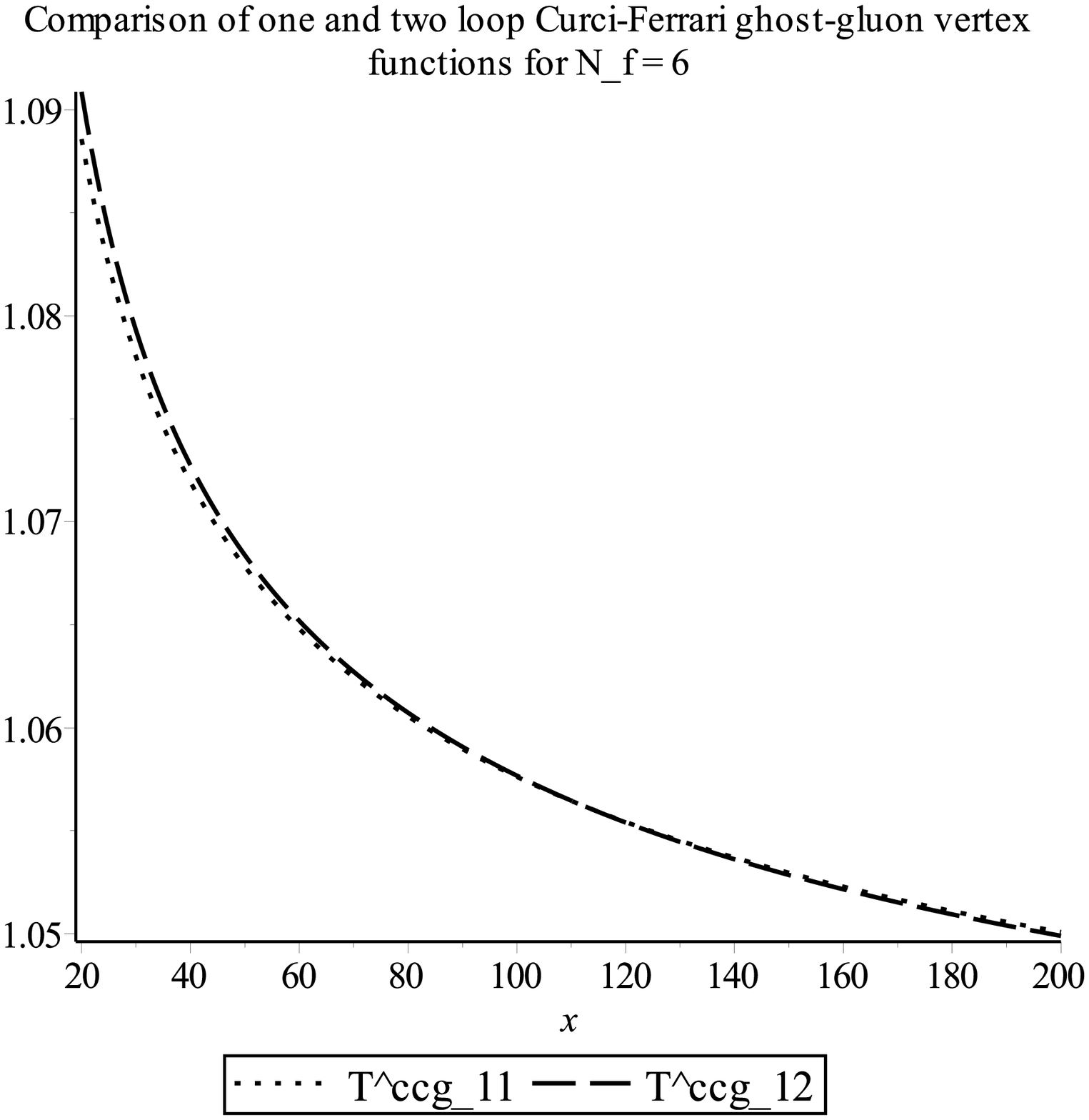}
\caption{Comparison of the $\MSbar$ ghost-gluon vertex function in the 
Curci-Ferrari gauge for different values of $\Nf$.}
\end{figure}}

The situation in the Curci-Ferrari gauge follows a similar pattern. However, as
the main difference in that gauge compared with the linear covariant gauge is
the nature of the ghost-gluon vertex we focus our discussion on the 
corresponding amplitudes. We have illustrated these in Figure $3$ for a range
of $\Nf$ in the $\MSbar$ scheme. Concerning the normalization we have chosen in
this case to plot the channel $1$ amplitude multiplied by a factor of $2$. This
is to allow one to compare with a similar plot for the linear covariant gauge 
given in \cite{69}. A similar picture emerges in that for larger values of 
$\Nf$ the one and two loop corrections are effectively the same. While the 
discrepancy looks large for smaller values of $\Nf$ at any specific value of
$x$ the variation is no more than $0.5\%$. If we compare the Curci-Ferrari 
gauge ghost-gluon vertices with the off-diagonal ghost-gluon vertex in the MAG
we see that at high momenta they are virtually indistinguishable. Where there 
is any difference it is at lower values of $x$. This is not unexpected as in 
effect at large energy the one loop piece of each vertex would be dominant.
Moreover, the one loop running of the coupling constant is both scheme and
gauge independent. 
 
As the $\MSbar$ results give an indication of the effect of the higher order
corrections and the small changes that the two loop contributions make, the
situation with the MOM schemes cannot be seen given that we are focused at the
symmetric point. Instead it seems appropriate to consider a physical quantity 
and compare values for it in the different schemes. In \cite{79} the flavour 
non-singlet $R$ ratio was evaluated in the MOM schemes of Celmaster and 
Gonsalves for the Landau gauge at three loops and compared with the $\MSbar$ 
scheme form, \cite{80,81,82,83,84,85,86}. Therefore, we have repeated that 
exercise for the MOM schemes of the MAG. First, we recall the notation used in
\cite{79} and define the $R$ ratio in scheme ${\cal S}$ by
\begin{equation}
R^{\cal S}(s) ~=~ \NF \left( \sum_f Q_f^2 \right) r^{\cal S}(s) ~.
\end{equation}
where $\NF$ is the dimension of the fundamental representation, $Q_f$ is the 
charge of the active number of quarks, $s$ is the centre of mass energy and
the perturbative expansion is defined by
\begin{equation}
r^{\cal S}(s) ~=~ \sum_{n=0}^\infty r_{n}^{\cal S}(s) {a^{\cal S}}^n
\end{equation}
and $r_0^{\cal S}$~$=$~$1$ in all schemes. From this the partial sums of the
series can be computed which are defined by
\begin{equation}
a_{pq}^{\cal S} \left( \frac{\mu^2}{{\Lambda^{\cal S}}^2} \right) ~=~
\sum_{n=1}^p r_n^{\cal S}(s)
\left( a_q^{\cal S}(\mu,\Lambda^{\cal S}) \right)^n ~.
\end{equation}
With these partial sums we have plotted 
$a_{22}^{\cal S} \left( \frac{\mu^2}{{\Lambda^{\cal S}}^2} \right)$ and
$a_{33}^{\cal S} \left( \frac{\mu^2}{{\Lambda^{\cal S}}^2} \right)$ for 
$\Nf$~$=$~$3$ and $5$ and presented representative results in Figure $4$. That 
we can analyse the two and three loop partial sums follows from the fact that 
we have the coupling constant maps from the MOM schemes to the $\MSbar$ ones at
two loops which allows us to construct the $R$ ratio at three loops. This is 
for the same reason why the three loop MOM $\beta$-functions can be constructed
in the MAG. In Figure $4$ we have included the $\MSbar$ result to compare with
and note that there is close agreement of the $\MOMq$ scheme with it. This is
not unexpected given that the $R$ ratio is based on a quark operator
correlation. As in \cite{79} the $\MOMg$ and $\MOMh$ scheme results lie further
away from the $\MSbar$ result due to the nature of the underlying quantity 
being considered in keeping with the original observations of \cite{38,39}. For
larger values of $\Nf$ there appears to be a larger discrepancy. However, while
this mimics the situation with the canonical linear covariant gauge, if 
anything the MOM schemes for the MAG lie closer to the $\MSbar$ result than the
former gauge. While the broadness of the estimate of the $R$ ratio at a 
particular centre of mass energy scale may appear large on the plot, the range 
is $5\%$ of a central value if one includes the $\MOMh$ scheme. While this may 
appear to be large the appropriate point is perhaps that this may be a better 
way of trying to estimate a theory error in a measurement in contrast to 
varying the actual running scale between two values chosen in an ad hoc manner.
What is also evident from these examples is that the specific value we have 
chosen in the MAG here, which is $\alpha$~$=$~$0$, is in keeping with the 
$\MSbar$ case which does not depend on the gauge parameter. For instance, in 
the linear covariant gauge the study of \cite{79} also illustrated that the 
Landau gauge versions of the $R$ ratio in the corresponding MOM schemes was 
consistent with $\MSbar$.  

{\begin{figure}[ht]
\includegraphics[width=7.6cm,height=8cm]{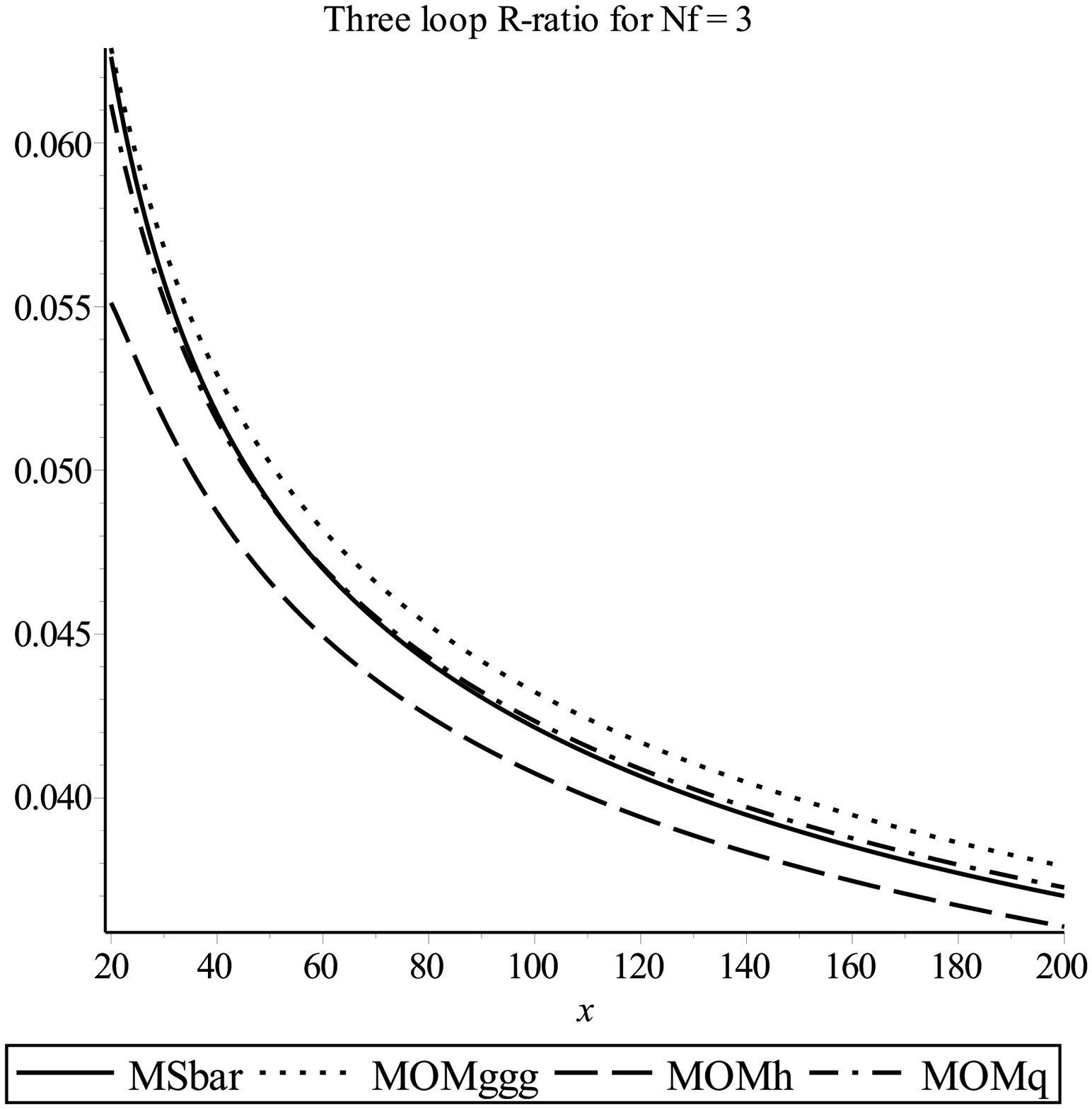}
\quad
\includegraphics[width=7.6cm,height=8cm]{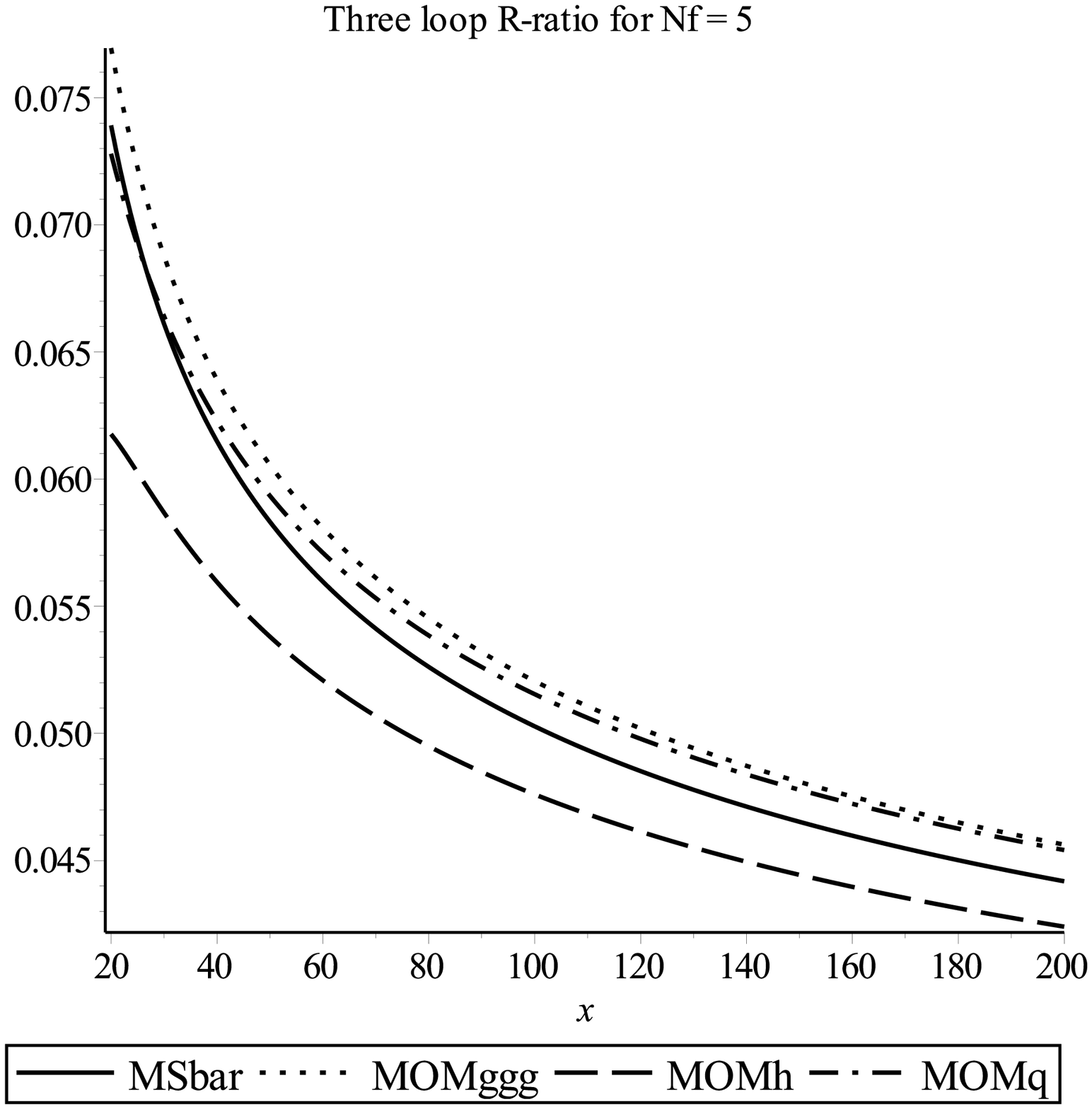}
\caption{Comparison of three loop $R$ ratio for $\Nf$~$=$~$3$ and $5$ in
various schemes.}
\end{figure}}

While we have considered the effect the various schemes have on a quantity of
experimental interest this was in the chiral limit. While this is an idealized
situation we make brief comments on the complexity of including quark mass
effects. First, at high energies quark masses can be neglected as a reasonable 
approximation. However, for lattice analyses where the matching is performed to
merge with the perturbative results such mass effects would be important at the
interface region short of the high energy limit. To estimate the errors on the 
massless vertex functions results by including physical masses is not 
immediately straightforward in the gauges considered here or in the linear 
covariant gauge. First, a full quark mass analysis would require all the 
relevant master integrals at one and two loops with propagators for all the
possible quark mass configurations. For two loops these are currently unknown. 
At one loop various masters are available, \cite{87,88}, and general results 
are known for the one loop triple gluon and quark-gluon vertices. For the 
ghost-gluon vertex the diagrams with a quark propagator do not appear until two
loops. Although the general results of \cite{87,88} provide a full analytic 
structure of the first of these two vertices the nature of the vertex functions
even at the symmetric point depend on several Clausen functions whose arguments
are ratios of the quark mass and $\mu^2$. However, a symmetric point analysis 
is too restrictive to quantify quark mass effects. Instead a more appropriate 
approach would be to compute the corrections to the fully off-shell vertex 
functions in powers of $m_q^2/\mu^2$ where $m_q$ is a generic quark mass. Such 
an analysis is well beyond the scope of the present article.

\sect{Discussion.}

The results presented in this article represent the completion of the programme
of studying QCD fixed in a variety of covariant gauges at two loops at the 
fully symmetric subtraction point. The one loop investigation for a linear 
covariant gauge was initiated several decades ago in \cite{38,39} which was 
extended to two loops in \cite{69}. In this article we have extended the one
loop MAG and Curci-Ferrari gauge analyses of \cite{37} to the same order as the
linear covariant gauge case of \cite{69}. In particular checks on the MAG 
results are inextricably entwined with those of the Curci-Ferrari gauge. 
Although nonlinear gauges are not necessarily the gauges of calculational 
choice for high energy analyses, the relation however, of the MAG to low energy
gluon and quark confinement, \cite{3,4,5,6,7,16}, suggests that for 
understanding mechanisms in this regime the MAG will be of analytic importance.
While the apparent difference in the ghost-gluon vertices in the MAG and 
Curci-Ferrari gauges is suggestive of such a picture, that observation is very 
much still within the perturbative regime. However, having the precision 
information on the vertex functions given here should ensure that 
Schwinger-Dyson models, and the assumptions behind the approximations made 
therein, have independent information to tally with. Ultimately the behaviour 
of Green's functions computed with Schwinger-Dyson techniques have to agree at 
high energy with perturbation theory. The two loop results will be useful in 
this respect. Although ultimately the next order will be of interest, that 
programme requires the determination of the three loop symmetric point master 
integrals in the Laporta approach.  

\vspace{1cm}
\noindent
{\bf Acknowledgements.} This work was carried out with the support in part
from STFC including a studentship (JMB) as well as through a John Lennon
Memorial Scholarship (JMB). JMB thanks Dr P. Dempster for useful discussions.
The {\sc Axodraw} package, \cite{89}, was used to draw Figure $1$.

\end{document}